\documentclass[structabstract]{aa}  

\usepackage{graphicx}
\usepackage{txfonts}
\usepackage[round]{natbib}
\bibpunct{(}{)}{;}{a}{}{,} 
\usepackage{epsfig}
\usepackage{longtable}

\begin{document}
   \title{Multi-wavelength study of the low-luminosity\\ outbursting
     young star HBC\,722\thanks{This work is based on observations
       made with the \emph{Spitzer} Space Telescope. \emph{Spitzer} is
       operated by the Jet Propulsion Laboratory, California Institute
       of Technology under a contract with NASA.}}

   \author{\'A. K\'osp\'al\inst{1}
           \and
           P. \'Abrah\'am\inst{1}
           \and 
           J. A. Acosta-Pulido\inst{2,3}
           \and
           M. M. Dunham\inst{4}
           \and
           D. Garc\'\i{}a-\'Alvarez\inst{2,3,5}
           \and
           M. R. Hogerheijde\inst{6}
           \and
           M. Kun\inst{1}
           \and
           A. Mo\'or\inst{1}
           \and
           A. Farkas\inst{1}
           \and
           G. Hajdu\inst{7,8}
           \and
           G. Hodos\'an\inst{9}
           \and
           T. Kov\'acs\inst{1}
           \and
           L. Kriskovics\inst{1}
           \and
           G. Marton\inst{1}
           \and
           L. Moln\'ar\inst{1}
           \and
           A. P\'al\inst{1,10}
           \and
           K. S\'arneczky\inst{1}
           \and
           \'A. S\'odor\inst{1}
           \and
           R. Szak\'ats\inst{1}
           \and
           T. Szalai\inst{11}
           \and
           E. Szegedi-Elek\inst{1}
           \and
           A. Szing\inst{12}
           \and
           I. T\'oth\inst{1}
           \and
           K. Vida\inst{1}
           \and
           J. Vink\'o\inst{1,11}}

   \institute{Konkoly Observatory, Research Centre for Astronomy and
     Earth Sciences, Hungarian Academy of Sciences, 
     PO Box 67, 1525 Budapest, Hungary\\
     \email{kospal@konkoly.hu}
     \and
     Instituto de Astrof\'\i{}sica de Canarias, Avenida V\'\i{}a L\'actea,
     38205 La Laguna, Tenerife, Spain
     \and
     Departamento de Astrof\'\i{}sica, Universidad de La Laguna,
     38205 La Laguna, Tenerife, Spain
     \and
     Harvard-Smithsonian Center for Astrophysics, 60 Garden Street, MS
     78, Cambridge, MA 02138, USA
     \and
     Grantecan S. A., Centro de Astrof\'\i{}sica de La Palma, Cuesta de San
     Jos\'e, E-38712 Bre\~na Baja, La Palma, Spain
     \and
     Leiden Observatory, Leiden University, Post Office Box 9513, 2300
     RA Leiden, The Netherlands
     \and
     Instituto de Astrof\'\i{}sica, Facultad de F\'\i{}sica,
     Pontificia Universidad Cat\'olica de Chile, Av. Vicu\~na Mackenna
     4860, Santiago, Chile
     \and
     Instituto Milenio de Astrof\'\i{}sica, Santiago, Chile
     0000-0003-0594-9138
     \and
     University of St Andrews, School of Physics \& Astronomy, St
     Andrews, United Kingdom 
     \and
     Department of Astronomy, Lor\'and E\"otv\"os University,
     P\'azm\'any P\'eter s\'et\'any 1/A, 1117 Budapest, Hungary
     \and
     Department of Optics and Quantum Electronics, University of
     Szeged, D\'om t\'er 9., 6720 Szeged, Hungary
     \and
     Baja Observatory, University of Szeged, 6500 Baja, KT: 766}
\date{Received date; accepted date}

 
 \abstract
  {HBC\,722 (V2493\,Cyg) is a young eruptive star in outburst since 2010. Spectroscopic evidences suggest that the source is an FU Orionis-type object, with an atypically low outburst luminosity.}
  {Because it was well characterized in the pre-outburst phase, HBC\,722 is one of the few FUors where we can learn about the physical changes and processes associated with the eruption, including the role of the circumstellar environment.}
  {We monitored the source in the $BVRIJHK_S$ bands from the ground, and at 3.6 and 4.5\,$\mu$m from space with the Spitzer Space Telescope. We analyzed the light curves and studied how the spectral energy distribution evolved by fitting a series of steady accretion disk models at many epochs covering the outburst. We also analyzed the spectral properties of the source based on our new optical and infrared spectra, comparing our line inventory with those published in the literature for other epochs. We also mapped HBC\,722 and its surroundings at millimeter wavelengths.}
  {From the light curve analysis we concluded that the first peak of the outburst in 2010 September was mainly due to an abrupt increase of the accretion rate in the innermost part of the system. This was followed after a few months by a long term process, when the brightening of the source was mainly due to a gradual increase of the accretion rate and the emitting area. Our new observations show that the source is currently in a constant ``plateau'' phase. We found that the optical spectrum was similar both in the first peak and the following periods, but around the peak the continuum was bluer and the H$\alpha$ profile changed significantly between 2012 and 2013. The source was not detected in the  millimeter continuum, but we discovered a flattened molecular gas structure with a diameter of 1700\,au and mass of 0.3\,M$_{\odot}$ centered on HBC~722.}
{While the first brightness peak could be interpreted as a rapid fall of piled-up material from the inner disk onto the star, the later monotonic flux rise suggests the outward expansion of a hot component according to the theory of Bell \& Lin (1994). Our study of HBC\,722 demonstrated that accretion-related outbursts can occur in young stellar objects even with very low mass disks, in the late Class\,II phase.}

\titlerunning{Multi-wavelength study of HBC\,722}
\authorrunning{K\'osp\'al et al.}

   \maketitle

\section{Introduction}

Sun-like pre-main sequence stars are surrounded by circumstellar
disks, from which material is accreted onto the growing protostar. The
accretion rate is variable: the protostar's normal accretion at a low
rate may be occasionally interspersed by brief episodes of highly
enhanced accretion \citep{kenyon1990}. FU Orionis-type variables
(FUors) are thought to be the visible examples of episodic
accretion. During their episodic ``outbursts'', accretion rate from
the circumstellar disk onto the star increases by several orders of
magnitude, from typically $10^{-7}$ up to
10$^{-4}$\,M$_{\odot}$\,yr$^{-1}$ \citep{audard2014}. Due to the
increased accretion, FUors brighten by 5--6\,mag at optical
wavelengths, their bolometric luminosities reach several hundred
L$_{\odot}$, and they stay in the high state for several decades.

Currently, about two dozen FUors and FUor candidates are known. One of
the recent discoveries, HBC\,722 went into eruption in 2010
\citep{semkov2010a}. The object's optical and near-infrared (near-IR)
spectra published by \citet{miller2011} and by \citet{semkov2012} are
similar to those of FU\,Ori and other FUors, confirming its FUor-type
classification. HBC\,722, however, differs from typical FUors in that
its outburst luminosity and accretion rate are only on the order of
$L_{\rm bol}$ = 10--20\,L$_{\odot}$, $\dot{M}$ =
10$^{-6}$\,M$_{\odot}$\,yr$^{-1}$, well below what is usual for FUors
\citep[][hereafter, Paper\,I]{kospal2011}. Despite this, HBC\,722 has
been in the bright state for at least five years now, it is currently
brighter than ever \citep{baek2015}, and it is on its way to exhibit a
typical, decades-long FUor-type light curve.

HBC\,722 is not an isolated object but a part of the LkH$\alpha$ 188
cluster, a group of optically visible young stars showing H$\alpha$
emission \citep{cohen1979}. Several Class 0/I embedded protostars, as
well as very young, very low luminosity protostars or starless cores
in the vicinity of HBC\,722 indicate active star formation in the area
\citep{green2011,dunham2012b}. HBC\,722 itself is a Class\,II object
with a circumstellar disk \citep{miller2011,kospal2011}, although with
a rather low upper limit for the disk mass \citep{dunham2012b}. The
system is surrounded by a reflection nebula \citep{miller2011}.

HBC\,722 is unique among FUors in that the progenitor, i.e., the
object in quiescence, has been well characterized. In order to learn
about the physical changes and processes associated to the outburst,
we carried out new optical and infrared photometric monitoring of
HBC\,722, including mid-IR observations, which provide new information
on the thermal emission of the inner disk at various phases of the
outburst. We also obtained optical and near-IR spectra, as well as
millimeter continuum and molecular line maps of the environment of the
source, and compared our results with the FUor outburst theory of
\citet{bell1994}.

\section{Observations and data reduction}
\label{sec:obs}

\subsection{Optical and infrared observations}

We obtained optical and near-IR images with $BVRIJHK_S$ filters
between 2010 September 19 and 2016 July 12 using four telescopes: the
Schmidt and RCC telescopes of the Konkoly Observatory (Hungary), as
well as the IAC-80 and TCS telescopes of the Teide Observatory in the
Canary Islands (Spain). Technical details of the telescopes and their
instrumentation are described in Paper\,I. Reduction of the images and
aperture photometry was performed in the same way as in Paper\,I. The
resulting magnitudes for the period between 2010 September 19 and 2011
January 2 are presented in Paper\,I, while the rest are listed in
Table~\ref{tab:phot}. The light curves are plotted in
Fig.~\ref{fig:light}.

We observed HBC\,722 using the \emph{Spitzer} Space Telescope in the
post-Helium phase at nine epochs between 2011 September 8 and 2012
October 12 (PID: 80165, PI: P.~\'Abrah\'am). We used the IRAC
instrument at 3.6 and 4.5$\,\mu$m in full array mode with exposure
time of 0.2\,s per frame. The data reduction and photometry was done
in the same way as described in detail in \citet{kun2011}, except that
here we used an aperture of 2 pixels (2$\farcs$4) and sky annulus
between 2--6 pixels (2$\farcs$4--7$\farcs$2). Convolution of the IRAC
filter profiles with the observed spectral energy distribution (SED)
proved that color correction is negligible. The results of the
photometry are listed in Table~\ref{tab:spitzer}.

We observed HBC\,722 with the LIRIS instrument installed on the 4.2\,m
William Herschel Telescope at the Observatorio del Roque de Los
Muchachos (Spain). The description of the instrument can be found in
\citet{acosta2007}. $JHK_S$ images and long-slit intermediate
resolution spectra in the $ZJ$ and $HK$ bands were obtained on 2011
July 20/21. The images were taken in a 5-point dither pattern, with
5\,s exposure time per dither position. The spectra were taken in an
ABBA nodding pattern with a total exposure time of 100\,s and 60\,s in
the $ZJ$ and $HK$ bands, respectively. We used the 0$\farcs$75 slit
width, which yielded a spectral resolution of R=550--700 in the
0.9--2.4$\,\mu$m range. We observed HIP\,103694, a G2-type star, as
telluric calibrator. The data reduction of both the images and the
spectra were done in the same way as in \citet{acosta2007}. The
spectra were flux calibrated using the $JHK_S$ photometry taken on the
same night. The typical signal-to-noise ratio of the spectra is
between 10 and 30. The photometry is included in Table~\ref{tab:phot},
while the spectra are plotted in Fig.~\ref{fig:liris}.

HBC\,722 was observed by the Wide-field Infrared Survey Explorer
(WISE; \citealt{wright2010}) in the cryogenic phase on 2010 May 28,
and in the framework of the NeoWISE program on 2010 November 25. For
both epochs, we downloaded all time resolved observations from the
AllWISE Multiepoch Photometry Table in the W1 (3.4\,$\mu$m) and W2
(4.6\,$\mu$m) photometric bands, and computed their average after
removing outlier data points. Since the beam sizes of WISE in the two
selected bands are approximately 6$''$, no contamination from
neighboring point sources is expected. We converted the computed
magnitudes to fluxes, and checked that similarly to Spitzer,
color-correction was unnecessary. In the errors, we added in
quadrature 2.4\% and 2.8\% as the uncertainty of the absolute
calibration in the W1 and W2 bands, respectively (Sect.~4.4 of the
WISE Explanatory Supplement). The resulting fluxes are listed in
Tab.\,2.

We carried out low-resolution spectroscopy of HBC\,722 with the
Optical System for Imaging and Low Resolution Integrated Spectroscopy
(OSIRIS) tunable imager and spectrograph \citep{Cepa03, Cepa10} at the
10.4\,m Gran Telescopio Canarias (GTC), located at the Observatorio
Roque de los Muchachos in La Palma, Canary Islands, Spain, on 2012
April 17 and on 2013 July 10. Observations were performed in service
mode within the GTC `filler' programs GTC55/12A and GTC3/13A. The
heart of OSIRIS is a mosaic of two 4k\,$\times$\,2k e2v CCD44--82
detectors that gives an unvignetted field of view of
7.8\,$\times$\,7.8\,arcmin$^{2}$ with a plate scale of
0.127\,arcsec\,pixel$^{-1}$. However, to increase the signal-to-noise
ratio (S/N) of our observations, we chose the standard operation mode
of the instrument, which is a 2\,$\times$\,2-binning mode with a
readout speed of 200\,kHz. All spectra were obtained with the OSIRIS
R2500R (red) grisms, covering the 5575--7685\,$\AA$ wavelength
range. Because of the highly variable seeing we used the 1$\farcs$23
slit oriented at the parallactic angle to minimize losses due to
atmospheric dispersion, providing a dispersion of 1.6 $\AA$/pixel. The
resulting wavelength resolution, measured on arc lines, was
$R$=2475. The exposure time was 60\,s in 2012 and 90\,s in 2013. The
spectra were reduced and analysed using standard IRAF routines.

\subsection{Millimeter observations}

We observed HBC\,722 at millimeter wavelengths with the Plateau de
Bure Interferometer (PdBI) and with the IRAM 30\,m telescope. The PdBI
observations were taken on two different nights, 2012 March 28 and
2012 April 2. The PdBI antennas were in 6Cq configuration, providing a
range of $uv$ radii between about 15\,m and 175\,m. The total
on-source correlation time was 2 hours. The weather conditions were
good, with precipitable water vapor between 2 and 8\,mm on March 28
and between 2 and 4\,mm on April 2. We used the 3\,mm receiver, tuned
it halfway between the $^{13}$CO and C$^{18}$O J=1--0 lines
(109.0918\,GHz), and placed a 20\,MHz-wide 39\,kHz-resolution
correlator unit on each CO line. We used two 160\,MHz-wide units as
well, in order to measure the 2.7\,mm continuum emission. At this
wavelength, the single dish HPBW is 45$\farcs$8. Bright quasars
(3C279, 2013+370, and J2120+445) were observed at regular intervals to
enable RF bandpass, phase, and amplitude calibration. The absolute flux
scale was fixed using the flux standard carbon star MWC\,349.  The
data was reduced in the standard way with CLIC, a GILDAS-based
application written especially for reducing PdBI data. The rms phase
noise was typically below 30$^{\circ}$. We estimate a flux calibration
accuracy of about 15\%.

The single-dish IRAM\,30\,m observations were taken during three
nights between 2012 June 19 and 22, with precipitable water vapor
between 5 and 9\,mm and stable weather conditions. We obtained
Nyquist-sampled 2$'{\times}$2$'$ on-the-fly (OTF) maps using the EMIR
receiver in frequency switching mode in the 110\,GHz band (HPBW was
22$''$). The map spacing was 7$''$ with 17 OTF subscans. One map was
obtained in about 7 minutes, and this script was repeated several
times to reach the desired sensitivity, in 6.25 hours. EMIR was used
with two backends simultaneously. The Versatile SPectrometer Array
(VESPA) provided 20\,kHz resolution and 60\,MHz bandwidth, which we
used to observe the $^{13}$CO and C$^{18}$O J=1--0 lines, to serve as
short spacings for the PdBI data. The Fast Fourier Transform
Spectrometer (FTS) was set in fine resolution mode, which we used to
cover a wide bandwidth of 4$\times$1.8\,GHz between 93\,GHz and
114\,GHz with a resolution of 50\,kHz, to look for additional lines in
the spectrum. The resulting maps have an rms noise of 0.04--0.05\,K.

\section{Results and analysis}
\label{sec:res}

\subsection{Light curves}
\label{sec:light}

\begin{figure*}
\centering \includegraphics[height=\textwidth,angle=90]{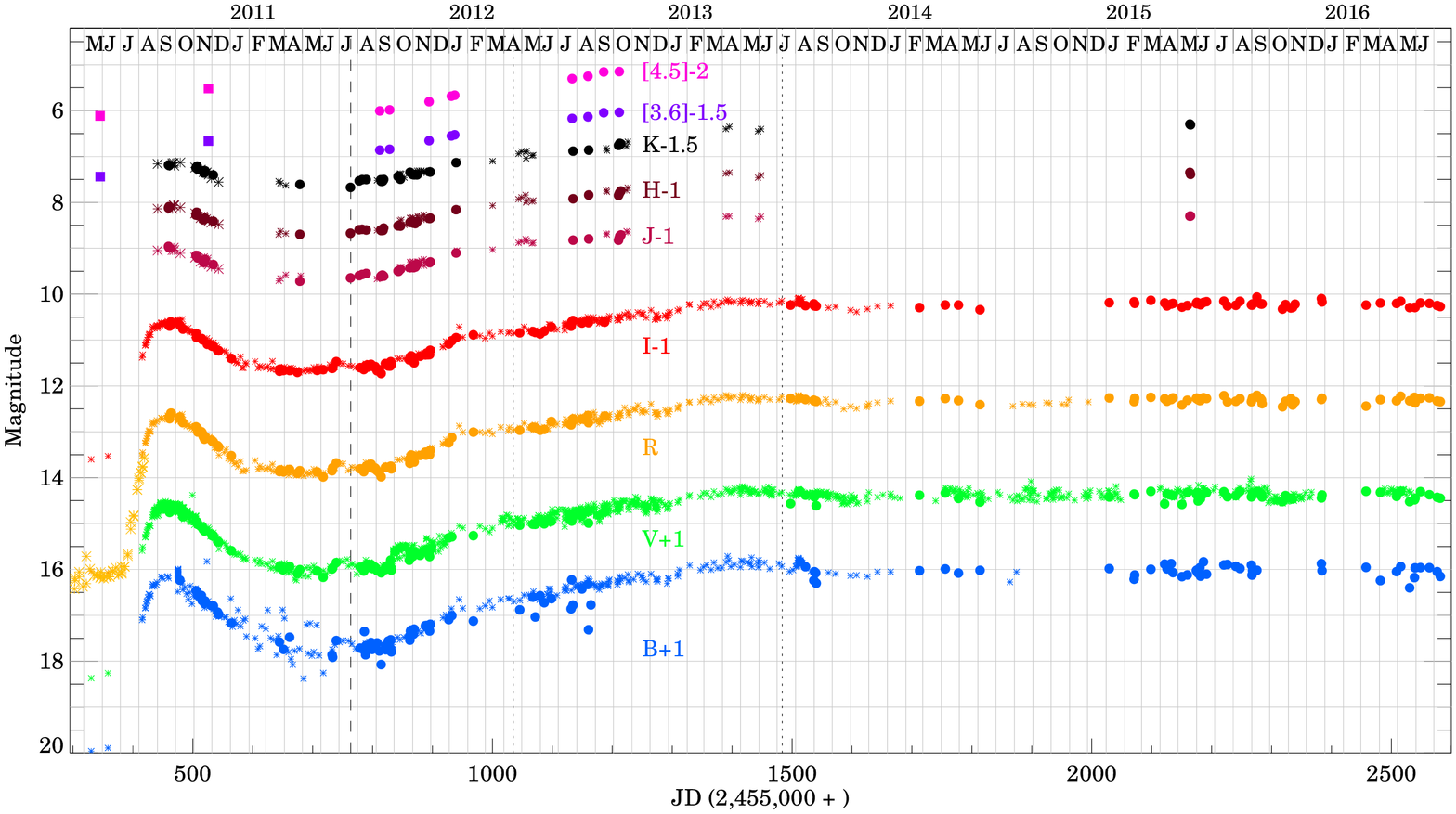}
\caption{Light curves of HBC\,722. Filled dots are from Paper I and
  this work, plus signs are from
  \citet{semkov2010,semkov2014,miller2011,sung2013,antoniucci2013},
  and from the AAVSO database (http://www.aavso.org). Mid-IR data
  points are either from WISE (filled squares) or from Spitzer (filled
  dots). For clarity, the $B$, $V$, $I$, $J$, $H$, $K_S$, [3.6] and
  [4.5] light curves are shifted along the y axis. Vertical dashed
  line marks the epoch when our near-IR LIRIS spectum was taken, while
  the vertical dotted lines indicates the time of our optical
  GTC/OSIRIS spectra.}
\label{fig:light}
\end{figure*}

Fig.~\ref{fig:light} displays the multiband light curves of HBC\,722,
compiled from our observations and literature data. In addition to
optical and near-IR measurements, our data set includes the
first dedicated mid-infrared (3.6 and 4.5\,$\mu$m) monitoring of a
one-year-long period of the outburst. The light curves show that the
rapid brightening of HBC\,722 started in 2010 mid-July, and peaked in
2010 September in all bands between $B$ and $K$
\citep{miller2011,semkov2014}. The maximum was followed by a slower
decay which lasted for about 5 months. Then the source stayed constant
for another 5 months, and started brightening again. This monotonic
flux rise could be clearly seen also in the Spitzer data points at 3.6
and 4.5$\,\mu$m. The increasing trend has a turning point in 2012
January, when the brightening slowed down. The fluxes at all
wavelengths reached a plateau in 2013 April, and our recent photometry
from 2015--2016 (Tab.~\ref{tab:phot}) demonstrates that they
have been approximately constant since then.

Optical monitoring of the source prior to the outburst
\citep{miller2011,semkov2012} revealed that what appears as the
starting point of the outburst in Fig.~\ref{fig:light} in 2010 May is
already an elevated flux level compared to the brightness of the
source before 2010. Thus, similarly to \citet{miller2011}, we will
consider the 2010 May state already part of the outburst, as an
initial pedestal where the rapid brightening started from (hereafter
``kickoff state'').

\begin{figure*}
\centering \includegraphics[width=\textwidth,angle=0]{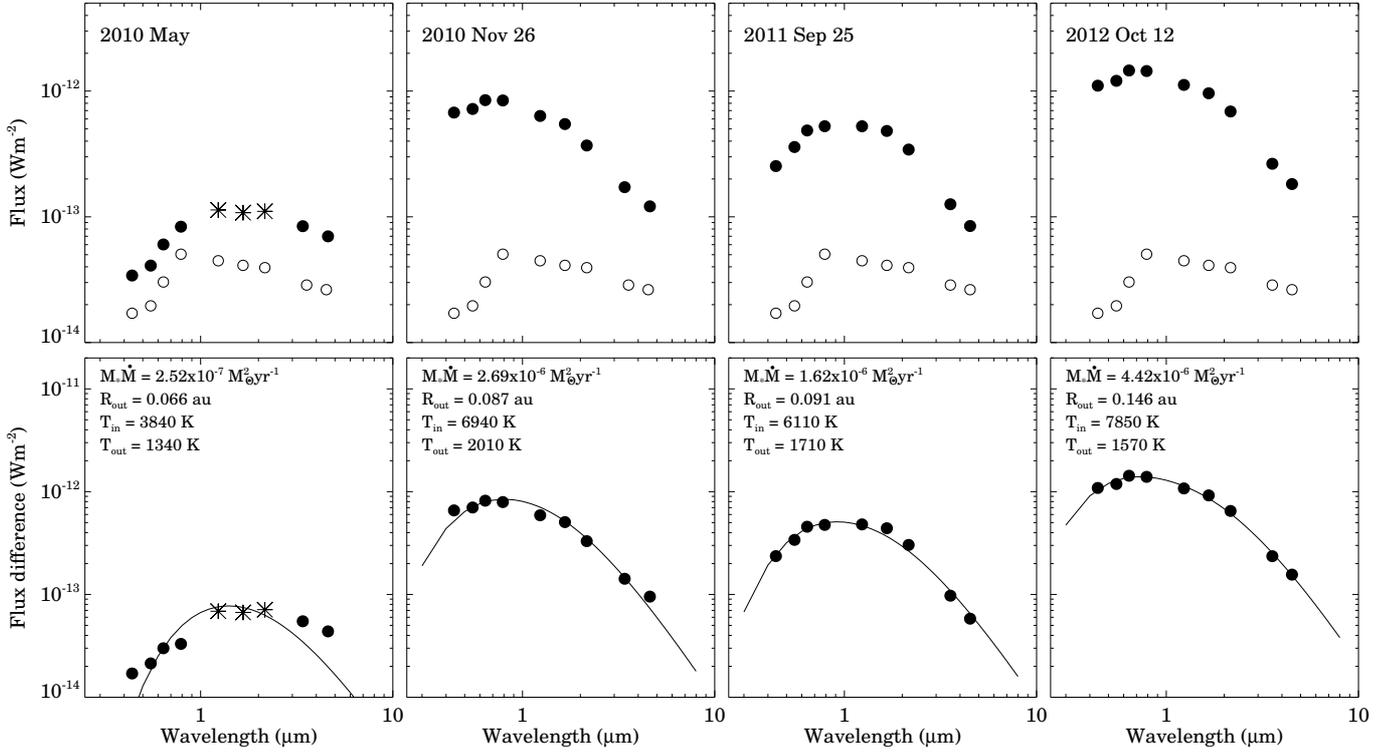}
\caption{SEDs of HBC\,722 at four different representative epochs of
  the outburst. Panels in the upper row present the actual SED (filled
  symbols), with the quiescent one from 2006 (open symbols)
  overplotted. The first epoch in 2010 May corresponds to the kickoff
  of the outburst just preceding the rapid first brightening, when
  the source was already in an elevated state compared to the
  quiescent one. In 2010 May the $JHK_S$ data points (asterisks) are
  extrapolations from later observations. The lower panels show the
  difference flux between the actual and the quiescent SEDs for the
  four epochs, as well as accretion disk fits to the difference.}
\label{fig:SED2p}
\end{figure*}

While the general shapes of the light curves in different bands are
similar, the amplitudes of the light variations in
Fig.~\ref{fig:light} are not the same. This wavelength dependence
carries information on the physical processes responsible for the flux
changes. We constructed a SED for each night when near- or
mid-IR observations were available. If simultaneous optical data
(obtained on the same night) were not available, we interpolated in
the $BVRI$ light curves. We created separate SEDs for the kickoff
state (2010 May) and for the true quiescent state (before 2010, not
plotted in Fig.~\ref{fig:light}) of the system. For the latter, we
collected photometric data obtained between 2006 March and October,
namely optical measurements from \citet{semkov2012}, UKIDSS $JHK_S$
from Paper\,I, and Spitzer 3.6 and 4.5$\,\mu$m data points from
\citet{rebull2011}. A comparison between the quiescent and the kickoff
SEDs can be seen in Fig.~\ref{fig:SED2p} (upper left panel).

The optical and near-to-mid infrared excesses arise from the innermost
part of the circumstellar disk, which we will model with an optically
thick, viscous accretion disk of radially constant mass accretion
rate. Our approach is similar to that of \citet{zhu2007} adopted for
FU\,Ori. The main difference is that we aim to reproduce only
broad-band photometry rather than line spectroscopy, thus a vertical
treatment of the disk atmosphere is not necessary. At each epoch we
will assume a steady, geometrically thin, optically thick disk, whose
radial temperature profile is \citep{hk96}:
\begin{equation}
T^4_d(R)=\frac{3GM{\dot M}}{8{\pi}R^{3}\sigma} \Bigg[1-\bigg(\frac{R_{*}}{R}\bigg)^{1/2}\Bigg]
\label{eqn:temp}
\end{equation}
where $T_d$ is the disk temperature at radius $R$, $M$ is the stellar
mass, $\dot M$ is the accretion rate, $R_{\rm *}$ is the stellar
radius, $G$ is the gravitational constant, and $\sigma$ is the
Stefan-Boltzmann constant.  In order to avoid the unphysical zero
temperature at $R=R_{*}$, we followed \citet{zhu2007} to prescribe
$T=T_{\rm max}$ within 1.36\,$R_{\rm *}$, the radius of maximum
temperature. Then we calculated the disk's SED by integrating the
fluxes of concentric annuli between $R_{*}$ and $R_{\rm out}$. The
annuli were assumed to emit blackbody radiation corresponding to the
temperature defined by Eqn.~\ref{eqn:temp}, at a distance of
550\,pc. The observed variations of the light curves can then be
interpreted in terms of temporal changes in two parameters: $M\dot M$
and $R_{\rm out}$.

We converted the observed magnitudes to fluxes using the appropriate
flux values for zero magnitudes from the literature. Since we intended
to analyze the extra brightening related to the outburst, we
subtracted the quiescent SED, which represents the sum of the
photosphere of the central star and the quiescent emission of the
(passive) circumstellar disk, from each epoch. Examples for four
representative epochs can be seen in Fig.~\ref{fig:SED2p}. Then, we
corrected for interstellar reddening by adopting $A_V$=3.1\,mag
\citep{miller2011}. We fitted our accretion disk model to the
resulting SEDs. At each epoch we used all available data points,
except the kickoff state (2010 May), when the $\lambda\geq3.5\,\mu$m
section of the SED markedly deviated from our accretion disk model and
was excluded from the fit. Accidentally low individual error bars
would result in extra weight on certain data points in the fitting
routine, influencing the fit result. Thus, we set a minimum formal
uncertainty value of 3\% for the fitting procedure, and all error bars
below this threshold were increased to 3\%.

For the stellar radius we adopted $R_{*}$=1.51\,$R_{\odot}$, computed
from a stellar luminosity of $L_*= 0.67$\,$L_{\odot}$, corresponding
to a Kurucz stellar photosphere with $T_{\rm eff}$ = 4250\,K at a
distance of 550\,pc \citep{kurucz2004}. This model atmosphere provided
the best match to our $BVRIJH$ data points in the quiescent phase. The
accretion disk model resulted in reasonable fits to the SEDs, however,
reproducing the shortest wavelength data points in the $B$- and
$V$-bands turned out to be sensitive to disk inclination. A closer to
edge-on geometry would correspond to smaller projected disk area,
which requires higher temperature, i.e., higher accretion rate, to
match the observed level of optical-infrared emission. A too high
temperature, however, would produce too much short wavelength
radiation, thus fitting the $B$- and $V$-bands offers a possibility to
constrain the inclination. Minimizing a total ${\chi}^2$, computed by
adding the ${\chi}^2$ values of the individual epochs for the whole
outburst, yielded an inclination value of 73$_{-15}^{+6}$\,deg, i.e.,
the system is seen in close to edge-on.

\begin{figure}
\centering \includegraphics[width=\columnwidth,angle=0]{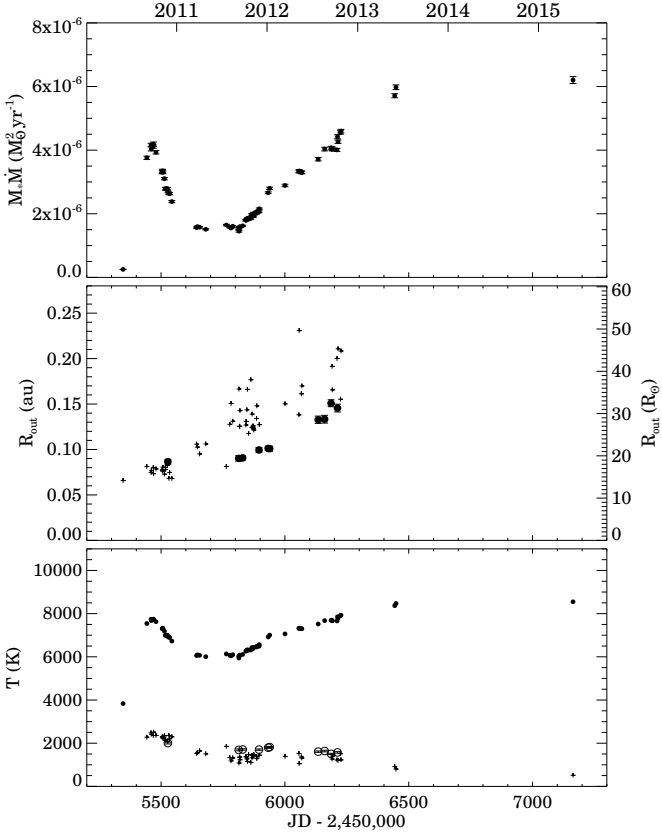}
\caption{Time evolution of the accretion rate (multiplied by the
  stellar mass), the outer radius, and the temperature at the inner
  and outer edges of the accretion disk. In the middle panel the black
  dots mark those epochs when $\lambda$$\geq$3.4\,$\mu$m data points
  are available for the fitting.}
\label{fig:fitpar2p}
\end{figure}

Adopting this inclination value, we obtained good fits for all SEDs in
our monitoring program. In Fig.~\ref{fig:SED2p} we overplotted the
best fit results at the four representative epochs. This demonstrates
that the excess emission related to the outburst is fully consistent
with an accretion disk profile. The time evolution of the fitted
parameters and derived temperature at the inner and outer edges of the
disk are plotted in Fig.~\ref{fig:fitpar2p}. Note that the outer
radius is better constrained when mid-IR data points are
available (black dots), while it has higher uncertainty at those
epochs when $K$-band was the longest wavelength included in the
fit. Nevertheless, all data points outline similar general trends.

Our simple accretion disk models indicate $M\dot M$ values of a few
times 10$^{-6}$\,$M_{\odot}^2$\,yr$^{-1}$. Assuming a stellar mass of
$0.8-0.9$\,$M_{\odot}$, typical for young K7-type stars (e.g.,
\citealt{siess2000}), the resulting accretion rates are somewhat
higher than published earlier for HBC\,722, calculated from the
accretion luminosity ($\sim$10$^{-6}$\,$M_{\odot}$\,yr$^{-1}$,
\citealt{kospal2011, green2013b}). The difference is mainly related to
the flat disk geometry and the more edge-on than face-on
inclination. Earlier results from the literature support the
relatively high accretion rate and close to edge-on
geometry. \citet{green2013} observed exceptionally strong [OI] line
emission at 63.18$\,\mu$m, suggesting stronger accretion than what is
calculated from the accretion luminosity. Our inclination angle is
close to the value of 85$^{\circ}$ suggested by \citet{gramajo2014}
from SED modelling. The rapid change in the line-of-sight extinction,
derived from multi-epoch X-ray observations by \citet{liebhart2014}
may also point to a more edge-on than pole-on configuration.

The uncertainties associated to the disk inclination, stellar mass,
distance, and especially the applicability of our simple disk model,
make the absolute value of the accretion rate uncertain. As a test, we
repeated the fitting process by relaxing the assumption in
Eqn.~\ref{eqn:temp} that the inner edge of the disk is equal to the
stellar radius, and let it be a free parameter in the fitting
process. While a trend for slightly increasing inner radius with time
was shown by the best fits, this model family requires an even closer
to edge-on geometry ($\geq$82$^{\circ}$), and 2--4 times higher
accretion rates, suggesting that the derived absolute values of $M\dot
M$ should be used with caution. However, the temporal trends seen in
the accretion rate and outer radius in Fig.~\ref{fig:fitpar2p} are
very robust, and we will analyze these trends in the following.

Figure~\ref{fig:fitpar2p} reveals that the first peak of the outburst
in 2010 September was caused by a rapid increase in the accretion
rate. This corresponded to a rise in the inner disk temperature. The
peak was followed by a large drop in the accretion rate, as well as a
somewhat smaller drop in the temperature, which explains the fading of
the source between 2010 September and 2011 February. During this
fading, the outer radius of the accretion disk stayed constant at
0.08--0.09\,au, with an outer disk temperature above 2000\,K. Next,
from 2011 autumn, the accretion rate started a gradual increase,
leading to higher maximum disk temperatures and causing the
re-brightening of the source. The outer radius also increased,
approximately linearly with time, while the temperature remained
constant at $\sim$1500\,K, due to the combined effects of increasing
accretion rate and increasing outer radius. The information on the
further evolution after 2012 November is sporadic and less conclusive,
because there are only three epochs with near-IR data, and no
mid-IR measurements. The derived accretion rate seems to
increase slightly further until mid-2013, then remains constant. The
fitted outer disk radii at these three epochs are unexpectedly large
(0.34\,au, 0.41\,au, and 0.77\,au; not plotted in
Fig.~\ref{fig:fitpar2p}), but also very uncertain. Nevertheless, the
formal outer disk temperatures at these three epochs became low, below
1000\,K.

Our results suggest that the outburst of HBC\,722 was driven by two
separate physical processes. The main one is the linear rise in the
accretion rate and the outer radius of the accretion disk that was
dominant from 2011 July until the plateau phase. Superimposed on this
is another process which caused the first brightness peak, dominated
the outburst evolution between 2010 August and 2011 February, and
disappeared afterwards.

\subsection{Optical and near-IR spectroscopy}

\begin{figure}
\centering \includegraphics[height=\columnwidth,angle=90]{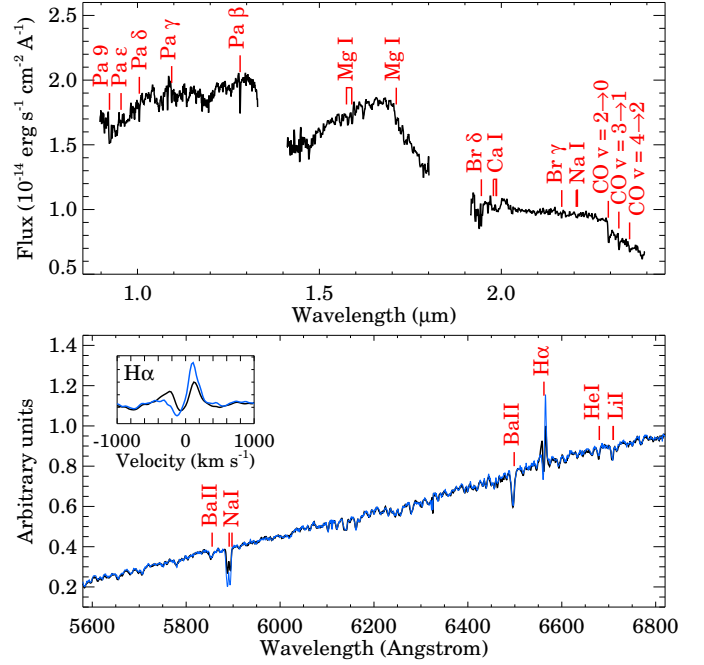}
\caption{{\it Top panel:} Near-IR spectrum of HBC\,722 obtained with
  WHT/LIRIS on 2011 July 21. To guide the eye, the vertical red bars
  show the laboratory wavelengths of atomic and molecular features
  typical for young stars. {\it Bottom panel:} Optical spectra of
  HBC\,722 obtained with GTC/OSIRIS on 2012 April 17 (blue curve) and
  2013 July 10 (black curve). The inset shows the H$\alpha$ line in
  more details as a function of the velocity. The identified lines are
  marked in red above the optical spectra.}
\label{fig:liris}
\end{figure}

Our near-IR spectrum in Fig.~\ref{fig:liris} exhibits several spectral
features, all in absorption. In the $ZJ$ band, Pa$\beta$ is the most
conspicuous line, but other lines of the Paschen series from
Pa$\gamma$ to Pa9 are also tentatively detected. The drop of the
spectrum towards longer wavelengths indicates the beginning of a broad
water band. In the $H$ band, no secure detection of lines is
possible. The general shape of the spectrum indicates another deep
water absorption band starting at 1.7$\,\mu$m. In the $K$ band, Ca
lines and the Br$\delta$ may be present. No detection of Br$\gamma$ is
evident. The CO band-head features are very well visible. Our spectrum
was taken after the first brightness peak, but before the source
started re-brightening (2012 July). \citet{lorenzetti2012} also
obtained near-IR spectra around the same phase of the light curve
(2012 August-September). Apart from the different spectral
resolutions, their and our spectra are similar, in the spectral
features, in the general spectral shape, and in the absolute flux
levels as well. All the features we detected were already present in
the spectrum taken during the first peak \citep{miller2011,
  lorenzetti2012}. The shape of our spectrum indicates redder emission
and more pronounced water absorption than at peak brightness. The
color change is consistent with the temperature changes we concluded
from our light curve analysis (Fig.~\ref{fig:fitpar2p}).

Our first optical spectrum was taken when HBC\,722 was
re-brightening. By the time we took the second optical spectrum, the
light curve reached the plateau phase. The spectra indicate gradually
rising continuum with the H$\alpha$ line displaying a complex profile,
and several other lines in absorption, including the Na D lines
(5892\AA, 5898\AA), two Ba\,II lines (6499\AA, 5855\AA), and the Li\,I
line (6709\AA). These lines are the same as those identified by
\citet{miller2011} in the corresponding wavelength range for a
spectrum taken close to the first brightness peak. Our two spectra are
remarkably similar, almost indistinguishable from each other, except
for the Na D lines, which became weaker, and for the H$\alpha$ line,
whose profile changed from 2012 to 2013 (Fig.~\ref{fig:liris}). The
H$\alpha$ line in the spectra of \citet{miller2011} showed a pure
emission feature. The monitoring of \citet{lee2015} revealed a time
evolution of the H$\alpha$ feature, which developed a P\,Cygni profile
already at the beginning of the outburst, attributed to a wind
component. Our spectrum from 2012 April also shows a clear P\,Cygni
profile. This is not unexpected, as a slowly evolving H$\alpha$
feature was observed for V1057\,Cyg \citep{herbig77} and for
V1647\,Ori \citep{aspin2009}. Interestingly, \citet{lee2015} found
that the wind component disappeared by the end of 2012, and then the
line profile exhibited a central absorption with emission wings on
both sides. Our measurement from 2013 July is consistent with this
line shape.

\subsection{Millimeter images}
\label{sec:mm}

\paragraph{Interferometric continuum data.}

\begin{figure}
\centering \includegraphics[width=\columnwidth,angle=0]{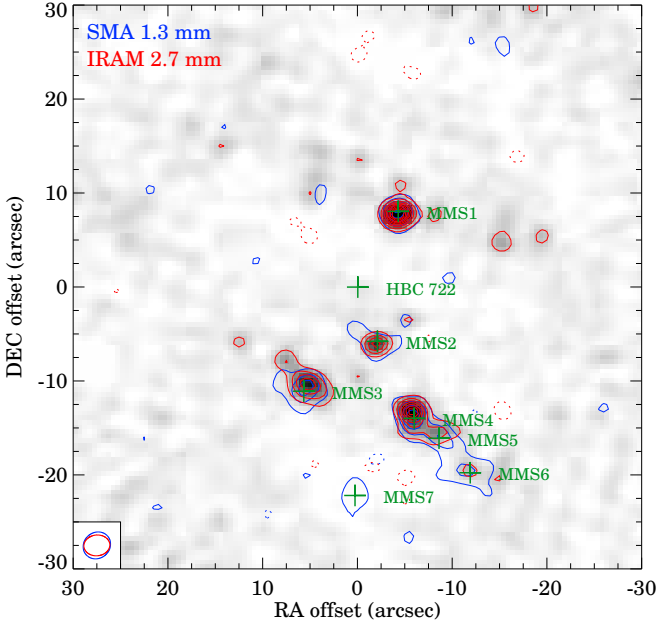}
\caption{Millimeter continuum emission map of the area around HBC
  722. The grayscale image and the red contours are IRAM PdBI 2.7\,mm
  observations from this work, while the blue contours are SMA 1.3\,mm
  data from \citet{dunham2012b}. At 2.7\,mm, contours are displayed at
  4, 8, 12, \dots $\sigma$, with $\sigma$ = 0.08\,mJy, while at
  1.3\,mm, contours are at 2.5, 5, 7.5, \dots $\sigma$, with $\sigma$
  = 1.65 mJy.}
\label{fig:mm}
\end{figure}

Fig.~\ref{fig:mm} shows our 2.7\,mm continuum image of the area around
HBC\,722. The target itself is not visible, and from the rms noise of
the image, we can give a 3$\sigma$ upper limit of
0.24\,mJy\,beam$^{-1}$ for its 2.7\,mm flux. This corresponds to a
point source upper limit for the total circumstellar mass of about
0.01\,M$_{\odot}$, using a gas-to-dust mass ratio of 100.
\citet{dunham2012b} mapped the area at 1.3\,mm with the SMA using
projected baselines between 5\,m and 76\,m
(3.8--58.5\,k$\lambda$). These largely overlap with the projected
baselines of our IRAM PdBI observations (between 15 and 175\,m or
5.6--64.8\,k$\lambda$), resulting in similar synthesized beams
(2$\farcs$73$\times$3$\farcs$02 at P.A.~$-$48$^{\circ}$ for SMA and
2$\farcs$74$\times$2$\farcs$21 at P.A.~96$^{\circ}$ for
IRAM). Therefore, the two images trace the same spatial scales and can
be safely compared. The comparison in Fig.~\ref{fig:mm} shows a
remarkable similarity between the two images at different millimeter
wavelengths. \citet{dunham2012b} did not detect HBC\,722 either, their
3$\sigma$ upper limit was 5\,mJy\,beam$^{-1}$ at 1.3\,mm, giving a
similar upper limit for the total mass as the value from our 2.7\,mm
measurement. They identified seven sources in the vicinity of
HBC\,722, named MMS1 to MMS7, marked in Fig.~\ref{fig:mm}. These
sources, claimed to be embedded protostars or starless cores by
\citet{dunham2012b}, may have formed together with HBC\,722. With the
exception of MMS7, we detected all these millimeter sources in our
2.7\,mm map. By fitting 2D Gaussians, we determined their peak fluxes,
total fluxes, deconvolved sizes, and position angles, which are listed
in Tab.~\ref{tab:iram}. Our photometry, together with that in
\citet{dunham2012b} enabled us to calculate the millimeter spectral
slopes of the detected sources. For this, we plotted the SEDs of the
sources $\nu F_{\nu}$ as a function of wavelength. For the three
sources with the best signal-to-noise millimeter detection, we
determined the slope $\alpha$ from the 1.3\,mm and 2.7\,mm points. For
optically thin emission, $\alpha$ is related to the spectral index of
the dust opacity, $\beta$, as follows: $\beta$ = $\alpha$ $-$ 3. For
the ISM, $\beta$ is about 1.7. If there is significant grain growth,
$\beta$ should be less, typically between 0 and 1
\citep[e.g.,][]{ricci2010}. We found that two sources, MMS3 and MMS4
have ISM-like dust ($\beta=1.6\pm0.3$ for MMS3 and $\beta=1.5\pm0.3$
for MMS4), while MMS1 shows signs of grain growth ($\beta=0.7\pm0.3$).

\paragraph{Interferometric and single-dish CO line data.}

\begin{figure*}
\centering \includegraphics[width=\columnwidth,angle=0]{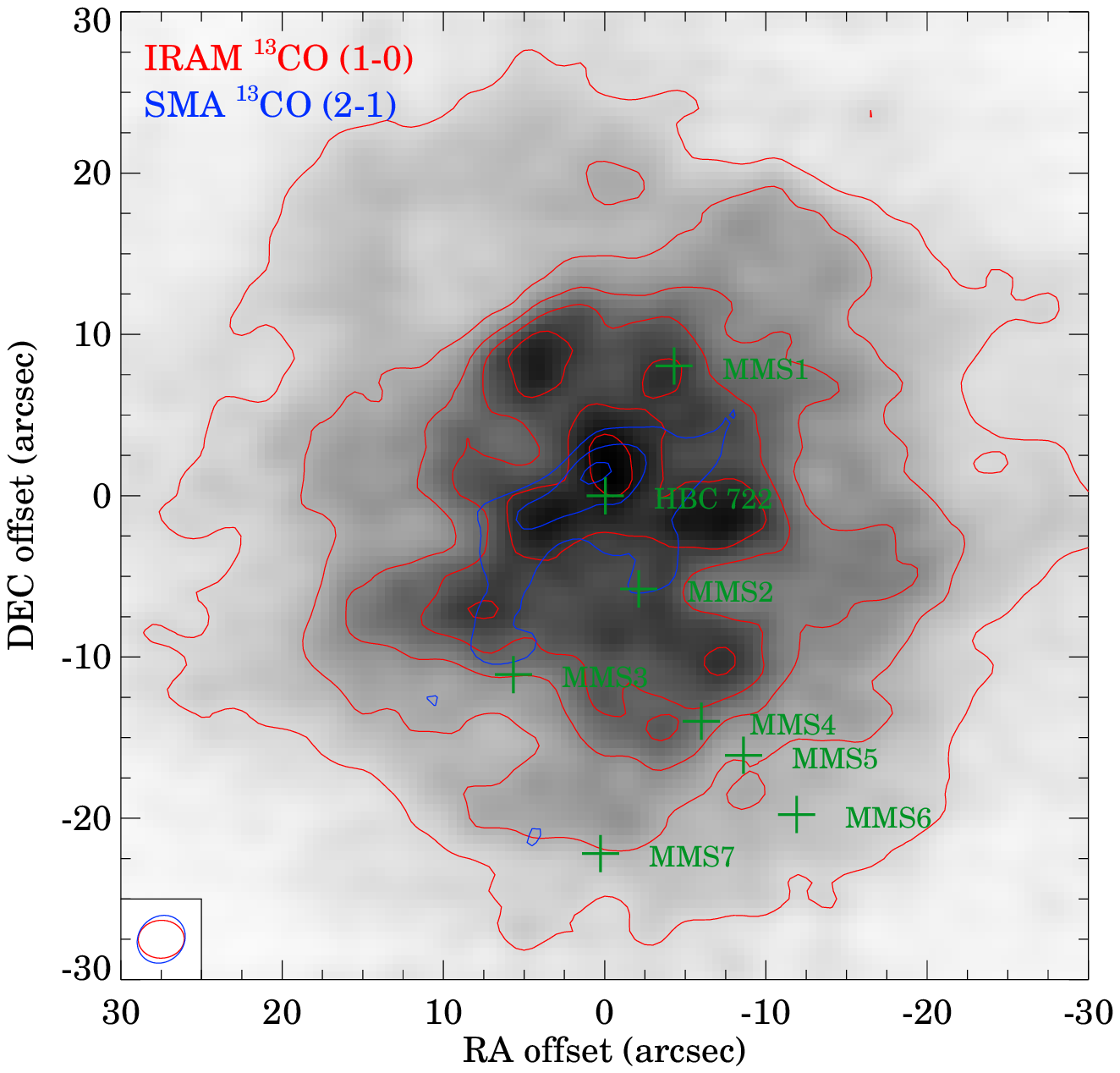}
\centering \includegraphics[width=\columnwidth,angle=0]{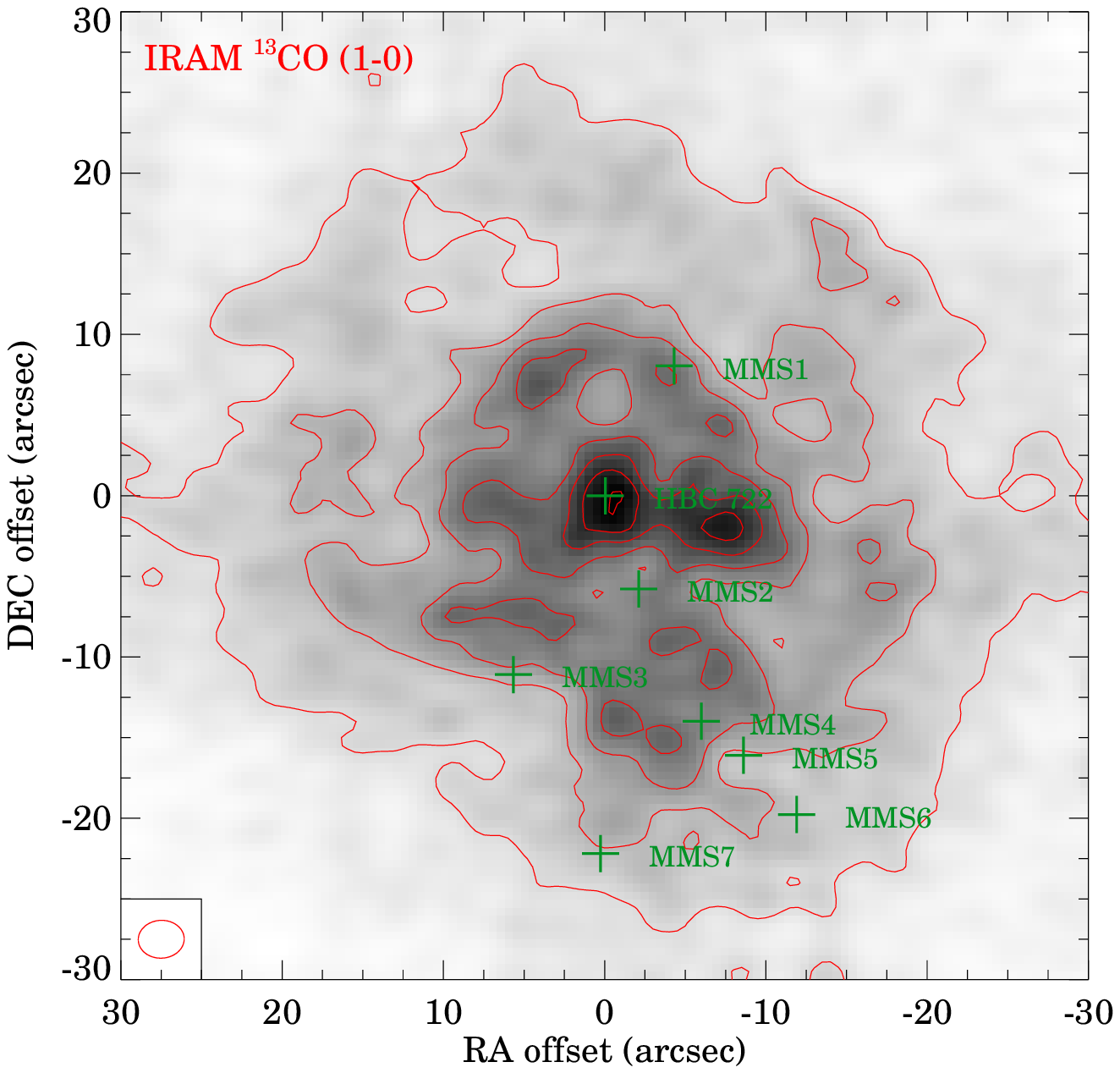}
\caption{{\it Left:} total intensity of the $^{13}$CO emission in the
  area around HBC~722, integrated for the 0--9\,km\,s$^{-1}$ velocity
  range. The grayscale image and the red contours are IRAM PdBI+30\,m
  observations from this work, while the blue contours are SMA data
  from \citet{dunham2012b}. For the $^{13}$CO(1--0) line, contours are
  displayed at 4, 8, 12, \dots $\sigma$, with $\sigma$ =
  0.09\,Jy\,km\,s$^{-1}$, while for the $^{13}$CO(2--1) line, contours
  are at 4, 8, 12, \dots $\sigma$, with $\sigma$ =
  0.3\,Jy\,km\,s$^{-1}$. {\it Right:} total intensity of the $^{13}$CO
  emission in the area around HBC~722, integrated for the
  5.46--6.52\,km\,s$^{-1}$ velocity range. Contours are at 4, 8, 12,
  \dots $\sigma$, with $\sigma$ = 0.03\,Jy\,km\,s$^{-1}$.}
\label{fig:mmb}
\end{figure*}

By examining the channel maps, we determined that significant CO
emission comes from the 0 -- 9\,km\,s$^{-1}$ velocity range. Figure
\ref{fig:mmb} {\it left} shows the total intensity of the
$^{13}$CO(1--0) emission integrated for this wide velocity range.
There is a roughly circular clump with a diameter of about 50$''$ with
several brightness peaks throughout the image, one of which may be
associated with MMS1, while the other compact mm sources are not
detected. Emission coinciding with the position of HBC~722 is detected
in a limited velocity range, between 5.45 and
6.52\,km\,s$^{-1}$. Fig.~\ref{fig:mmb} {\it right} shows the total
intensity of the $^{13}$CO(1--0) emission integrated for this narrow
velocity range. The structure centered on HBC~722 is slightly
elongated in the northwest--southeast direction with a deconvolved
size of about 1700\,au, while it remains unresolved in the
perpendicular direction. The integrated flux of this source is
1.7\,Jy\,km\,s$^{-1}$, corresponding to a total gas mass of
0.03\,M$_{\odot}$ assuming a temparature of 32\,K (see below). The
source is also visible in the C$^{18}$O map integrated for the same
narrow velocity range, and its flux of 0.22\,Jy\,km\,s$^{-1}$ gives
the same gas mass estimate, indicating that both lines are optically
thin.

In Fig.~\ref{fig:mmb} {\it left}, for reference, we also indicated
with contours the $^{13}$CO(2--1) emission measured with SMA by
\citet{dunham2012b}, integrated for the wide 0--9\,km\,s$^{-1}$
velocity range. Note that the two images cannot be directly compared,
because our IRAM map combines the interferometric and single dish
data, thus includes zero and short spacings, while the SMA map only
contains interferometric measurements, thus some extended flux is
filtered out. Spatial filtering probably explains the significant
differences between the IRAM and SMA maps. In fact, if we restrict the
imaging to the PdBI data and do not include short spacings from the
IRAM 30\,m telescope, we obtain a similar map to that of
\citet{dunham2012b}. Assuming local thermodynamic equilibrium (LTE)
and optically thin emission for the $^{13}$CO isotopologue, the two
interferometric-only maps can be used to estimate the temperature of
the emitting gas according to the Maxwell-Boltzmann distribution. For
this purpose we integrated the flux in an area of 10$'\times$10$'$,
20$'\times$20$'$, and 40$'\times$40$'$ centered on HBC\,722. We
obtained temperatures of 32\,K, 22\,K, and 20\,K, respectively, thus,
there is a hint for a temperature gradient.

With IRAM, we also obtained a map of the C$^{18}$O(1--0) emission,
which shows a spatial distribution very similar to that of
$^{13}$CO(1--0). The line ratio of the two isotopologues is around
8--9 throughout the map. The standard $^{13}$CO/C$^{18}$O abundance
ratio in the ISM is 8.1 \citep[e.g.,][]{wilson1994}, therefore our
measurements indicate that both C$^{18}$O and $^{13}$CO(1--0) are
optically thin, and can be used to estimate the total gas mass in the
area. Using 22\,K as a representative temperature, and the canonical
H$_2$/$^{12}$CO abundance ratio of 10$^4$, $^{12}$CO/$^{13}$CO abundance
ratio of 69, and $^{12}$CO/C$^{18}$O abundance ratio of 560
\citep{wilson1994}, we obtained 7.6\,M$_{\odot}$ from the $^{13}$CO
flux, and 7.3\,M$_{\odot}$ from the C$^{18}$O flux. The calculated
mass is between 6.9\,M$_{\odot}$ and 10.1\,M$_{\odot}$ for
temperatures between 20\,K and 32\,K.

We calculated first and second moment maps from our IRAM data
cubes. We detected no significant velocity gradient in the area of
HBC\,722. The line profiles are fairly similar at all points of the
maps with an average systemic velocity of
4.7\,$\pm$\,0.2\,km\,s$^{-1}$. The line profiles are fairly broad, the
average line widths are 3.4\,$\pm$\,0.3\,km\,s$^{-1}$ for $^{13}$CO
and 3.1\,$\pm$\,0.4\,km\,s$^{-1}$ for C$^{18}$O.

The size of the CO emitting region in Fig.~\ref{fig:mmb} {\it left} is
rougly consistent with the extension of the optical reflection nebula
visible around HBC~722 in outburst \citep{miller2011}. However, while
the CO emission is rather spherical, the optical nebula is highly
asymmetric, being more extended towards the southwest, indicating
anisotropic illumination. \citet{armond2011} observed that HBC~722 is
surrounded by H$\alpha$, [SII], and H$_2$ knots within 10$''$ of the
star that they call HH~655. There is no clear CO counterparts of these
structures in our CO channel maps. We searched for outflow signatures
as well, but detected none.

\begin{figure*}
\centering
\includegraphics[height=85mm,angle=90]{hbc_30m_5.ps}
\hspace*{5mm}
\includegraphics[height=85mm,angle=90]{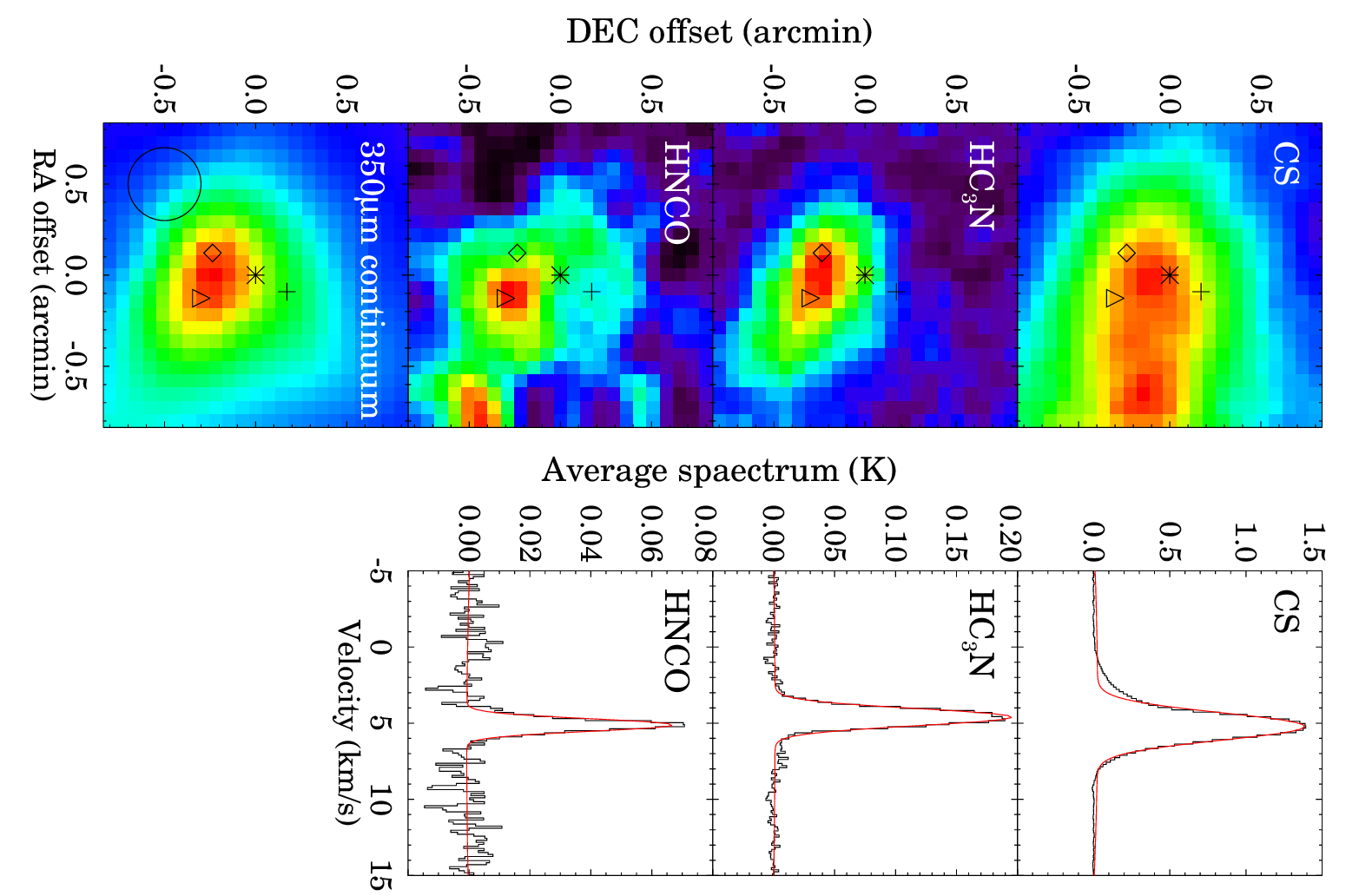}
\caption{{\it Left:} single dish IRAM 30\,m images of the area around
  HBC 722 in different lines, and a Herschel/SPIRE 350$\,\mu$m
  continuum image of the same area. For the IRAM data, the colors show
  the integrated line intensity and the scale goes from 0 to $I_{\rm
    max}$, where $I_{\rm max}$ is 25.0, 2.9, 1.4, 3.1, 6.2, 1.4, and
  0.25 K\,km\,s$^{-1}$ for the $^{13}$CO J=1--0, C$^{18}$O J=1--0,
  C$^{17}$O J=1--0, CN N=1--0, CS J=2--1, HC$_3$N J=12--11, and HNCO
  J$_{\rm K_{-1} K_{+1}}$=5$_{05}$--4$_{04}$ lines, respectively. For
  the Herschel image, the scale goes from 0 to 1050\,MJy\,sr$^{-1}$.
  The asterisk, plus, diamond, and triangle mark the positions of HBC
  722, MMS1, MMS3, and MMS4, respectively. Circles in the bottom left
  corners of the top and bottom panels mark the beam sizes of the
  IRAM\,30\,m telescope and Herschel. {\it Right:} Profiles of the
  lines averaged over the areas displayed in the left column (black
  histograms), with fitted Gaussian functions (red curves).}
\label{fig:30m}
\end{figure*}

\paragraph{Single-dish line data.} In order to search for emission
from different molecules, we obtained a wide bandwidth spectrum with
the FTS spectrograph on the IRAM\,30\,m telescope. We identified the
following species and transitions (in order of increasing frequency):
\begin{itemize}
\item the CS J=2--1 line at 97.981\,GHz,
\item the HC$_3$N J=12--11 line at 109.174\,GHz,
\item the C$^{18}$O J=1--0 line at 109.782\,GHz,
\item the HNCO J$_{\rm K_{-1} K_{+1}}$=5$_{05}$--4$_{04}$ line at 109.906\,GHz,
\item the $^{13}$CO J=1--0 line at 110.201\,GHz,
\item the C$^{17}$O J=1--0 hiperfine triplet at 112.359\,GHz, and
\item several CN N=1--0 lines at 113.123, 113.144, 113.170, 113.191, 113.488,
113.491, 113.500, 113.509, and 113.520\,GHz.
\end{itemize}
These are all low excitation transitions with upper level energies of
5.3\,K for $^{13}$CO and C$^{18}$O, 5.4\,K for C$^{17}$O and CN,
7.1\,K for CS, 10.6\,K for HNCO, and 34.1\,K for HC$_3$N. The line
profiles, obtained by integrating the data cubes for an area of
100$''\times$100$''$, are plotted in the right column of
Fig.~\ref{fig:30m}. The line profiles are approximately Gaussian, with
FWHM between 0.6\,km\,s$^{-1}$ and 2.2\,km\,s$^{-1}$ ($^{13}$CO is the
widest, HNCO is the narrowest), and the lines peak at a velocity
between 4.6\,km\,s$^{-1}$ and 6.0\,km\,s$^{-1}$.

The left column of Fig.~\ref{fig:30m} displays the total line
intensity maps integrated between --2\,km\,s$^{-1}$ and
10\,km\,s$^{-1}$. The CO emission forms a large round clump, similar
in all three isotopologues. Other lines are more compact, peaking in
the triangle formed by HBC\,722, MMS3, and MMS4. The CS emission is
more elongated towards the west. The matching velocities, similar
spatial locations, and low excitation temperatures indicative of cold
gas suggest that all gas lines are probably coming from the cloud that
formed the young stars and cores.

For comparison, we plotted in Fig.~\ref{fig:30m} the Herschel/SPIRE
350\,$\mu$m image of the area around HBC\,722. At this wavelength,
Herschel's beam size is approximately the same as that of the IRAM
30\,m telescope at 3\,mm. The spatial distribution of 350$\,\mu$m
continuum emission is very similar to that of the CN and HC$_3$N
lines, and to a lesser degree also to the CS and HNCO lines. It peaks
around MMS3 and MMS4, and is relatively compact. The emission of the
three CO isotopologues, on the other hand, is more extended. This is
not surprising, as the critical density is 2 --
3$\times$10$^3$\,cm$^{-3}$ for CO and 4$\times$10$^5$ --
2$\times$10$^6$\,cm$^{-3}$ for the other molecules, which are high
density gas tracers (e.g., \citealt{lequeux2005,gratier2013}, see also
the Leiden Atomic and Molecular
Database\footnote{http://home.strw.leidenuniv.nl/\~{}moldata/}). Thus,
CN, CS, NC$_3$N, and HNCO emission only comes from the densest part of
the cloud, while CO is emitted throughout the area. Dust continuum
emission, as evidenced by the SPIRE images, also traces the densest
part of the cloud.

\section{Discussion}
\label{sec:dis}

Based on its SED, HBC\,722 used to be a normal T\,Tauri-type star in
quiescence, except for a flux excess in the 2.2--8\,$\mu$m wavelength
range compared to the Taurus median \citep{kospal2011}. This excess
indicates a stronger than usual thermal emission from the inner disk,
pointing to either a large illuminated surface area (strong disk
flaring, inner envelope), or internal disk heating due to
accretion. The fact that the excess is present already at wavelengths
as short as the $K$-band suggests that the dust disk probably extended
inward to the sublimation radius (0.06--0.07\,au for a K7--M0 star).
The optical and near-IR SED of the quiescent system was consistent
with a reddened K7-type stellar photosphere shortward of the $K$-band
(\citealt{miller2011}, see also Sect.\,\ref{sec:light}), which shows
that the accreting gas within the dust sublimation radius, if present,
did not contribute significantly to the optical emission, and formed
an optically thin inner gas disk in quiescence.

We compared the quiescent SED (constructed in Sect. 3.1,
representative of the source in 2006 March-October) with photometric
observations obtained in 2010 May, just preceeding the rapid
brightening of the source (\citealt{semkov2010}, \citealt{miller2011},
WISE, this work). Our analysis revealed that HBC\,722 in this kickoff
state was already significantly brighter than in quiescence. The light
curves published by \citet{semkov2010} and \citet{miller2011} show
that this gradual flux rise started about six months before the
outburst. The brightening had similar amplitudes from $B$ to $R$, a
possible minimum in the $I$-band, and larger amplitudes again from the
$J$-band to 4.5$\,\mu$m, almost 1.3 times higher than in the $V$-band.

Based on these colors, the optical brightening in the kick-off state
is consistent with the appearance of a new component with a
temperature similar to the stellar temperature. If fitted with the
accretion disk model of Eqn.~\ref{eqn:temp} (Fig.~\ref{fig:SED2p},
bottom left panel), the temperature at the inner edge of the disk was
3830\,K, powered by an accretion rate of $M\dot M = 2.5
\times$10$^{-7}$\,$M_{\odot}^2$\,yr$^{-1}$
(Fig.~\ref{fig:fitpar2p}). The outer radius of the fitted accretion
disk is $\sim$0.07\,au, thus it resided within the sublimation radius
of the dust disk. The mid-IR data points are higher than the fitted
accretion disk model, suggesting increased thermal dust emission,
possibly due to dust structures situated above the accretion disk
surface (e.g., a vertical inner disk wall) at the dust sublimation
radius, in accordance with its temperature. At later epochs, this
excess is not visible and the accretion disk model reproduces the
observed SEDs, indicating that these dust structures
disappeared. Thus, it is tempting to speculate that in the kickoff
state we might witness the material that is being accumulated before
the outburst close to the sublimation radius, and which later fell
onto the star during the eruption. An event similar to the several
months long ``preparatory phase'' of the HBC\,722 outburst may be seen
in the light curve of the latest FUor V960\,Mon \citep[Figure 1
  in][]{hackstein2015}. If pre-outburst flux rise turns out to be a
general feature of FUor outbursts, light curves obtained by all-sky
surveys could be used to discover imminent FUor outbursts.

As described before in Section~\ref{sec:light}, the first brightness
peak in the light curves was due to a short maximum in the accretion
rate. Apparently, a packet of material arrived onto the star. The mass
of the infalling material, estimated by integrating the accretion rate
curve in Fig.~\ref{fig:fitpar2p}, was
$\sim$1.8$\times$10$^{-6}$\,M$_{\odot}$. Interestingly, the accretion
disk that reproduced the SED in the second column in
Fig.~\ref{fig:SED2p} has an outer radius on the order of the dust
sublimation radius. If we follow our speculation on the material that
accumulated in the inner dust disk before the outburst, then the first
peak in the light curves may be the consequence of the infall of the
accumulated material onto the star. Indeed, the freefall timescale at
the dust sublimation radius is several days. Therefore it is plausible
that this mass reservoire at the sublimation radius was emptied in a
few weeks. The appearance of a short brightness maximum at the very
beginning of the eruption is a characteristic feature of an outburst
model family proposed by \citet{bell1995}. They postulate that
eruptions are due to a thermal instability in the disk, and if the
instability is triggered farther from the star than the inner edge of
the disk, an ionization front, separating the cooler metastable and
the hot high state areas, will propagate rapidly toward the star. Such
initial peak was observed, e.g., in V1057~Cyg. In the case of HBC~722
the trigger could have occurred around the inner edge of the dust
disk, where matter was piled up before the outburst.

During the following 5 months, between 2011 April and August, the
brightness of HBC\,722 stayed more or less constant. The fact that
this flat part in the light curve started first in the infrared, where
the first peak was weaker and appeared only later in the B-band where
the first peak was the strongests suggests that here we simply see the
overlap between two parts of the light curve: the decaying first peak
and the emerging second brightness increase, which resulted in a
transition period of approximately constant brightness. The light
curve suggests that the process causing the first peak completely
disappeared by mid-2011, and it is unclear from our data whether this
process plays an essential role in the outburst mechanism.

The brightness rise between 2011 August and 2013 April is the most
interesting part of the outburst. According to
Fig.~\ref{fig:fitpar2p}, the disk area participating in the outburst
is expanding as its outer radius grows approximately linearly in time,
from $\sim$0.07\,au to $\sim$0.15\,au. This outward propagation
resembles the predictions of the thermal instability model of
\citet{bell1994}, where a hot, partially or fully ionized inner disk
expands into the colder outer material, triggering a thermal
instability at the ionization front that separates the two
phases. Since this second ionization front propagates outward, its
pace is significantly lower than the one during the first peak (see
above). The linear increase of the radius of the accretion disk
supports this hypothesis. The expansion of an ionization front stops
at an equilibrium radius, where the ionization and recombination rates
are equal, just like our data indicates. The expansion velocity of the
outer radius of the accretion disk in Fig.~\ref{fig:fitpar2p} measured
between 2011 August and 2013 April is on the order of
240\,m\,s$^{-1}$. This value is higher than the speed computed for a
standard model of \citet{bell1994} for an outward-propagating
ionization front, determined by the time sufficient for mass transfer
from the ionized region outwards to increase the surface density
enough for the thermal instability.

According to our modeling, the outward expanding accretion disk
extends over the sublimation radius of 0.07--0.08\,au, thus the
accretion disk may overlap with the dust disk. In the plateau phase,
after 2013 April, the accretion rate reached a level even higher than
the peak in 2010 September, and is has been constant for several
years. The total accreted mass in this phase is significantly higher
than in the first peak. We predict that the plateau phase will end
when the matter available in the inner disk depletes. Assuming a
simple disk model with a power law surface density density
distribution of $\Sigma \sim r^{-1.5}$, and a total disk mass of
0.01\,M$_{\odot}$ (Sect.\,\ref{sec:mm}), the innermost 0.15\,au region
contains enough mass to maintain the outburst for about 18
years. After the depletion of the inner disk the outburst stops,
possibly leaving an extended, optically thin inner hole in the disk,
which may be noticeable in the SED as a lack of mid-IR
excess. Considering that we saw no indication of an inner hole in
HBC~722 before the outburst, this hole should probably fill up before
the system can erupt again.

The accretion rate changes in the HBC~722 system during its outburst,
as outlined by our measurements, can be compared with the conclusions
drawn from multi-epoch X-ray observations by
\citet{liebhart2014}. They obtained three X-ray measurements, the
first one during the first peak of the light curve, the second one
during the local brightness minimum before the re-brightening, and the
third one during the plateau phase. They suggest that the first peak
was due to an initial strong disk instability that rapidly led to
enhanced accretion. It transported a huge amount of dust-free gas to
the vicinity of the star, which absorbed all X-ray emission from the
innermost region. This is consistent with our interpretation of the
first brightness peak in the light curves in terms of a short maximum
in the accretion rate. At their second X-ray epoch,
\citet{liebhart2014} could marginally detect the star, indicating
significantly less gas around the star, in accordance with our picture
where the first peak was followed by a rapid drop of the accretion
rate. Finally, \citet{liebhart2014} proposed that by their third epoch
the accretion rate increased again, producing enhanced absorption. Our
Fig.~\ref{fig:fitpar2p} shows that the accretion rate indeed increased
by a factor of four between the dates of the second and third X-ray
measurements.

If the present accretion rate remains, HBC~722 would accrete about
1$\times$10$^{-4}$\,M$_{\odot}$ during the hypothesized length of the
present outburst, 18 years. The disk mass, however, is relatively low,
below 0.01\,M$_{\odot}$, which allows not more than 100 such eruptions
to consume the whole circumstellar disk. A possible source of
replenishment of material could be the flattened envelope of
0.03\,M$_{\odot}$, detected in molecular gas emission in
Sect.\,\ref{sec:mm}. A detailed investigation of this structure,
including the exploration of its infall pattern and mass infall rate,
is indispensable to answer this question.

\section{Summary and Conclusions}
\label{sec:con}

Based on our optical-infrared monitoring, spectroscopic observations,
and millimeter mapping, we performed a detailed study of the first six
years of the outburst of the low-mass T\,Tauri star HBC\,722. We
interpreted the first brightness peak, lasting several months, as the
rapid fall of piled-up material from the inner disk onto the
star. This was followed by a monotonic flux rise, which can be
explained by increasing accretion rate and emitting area. Our
observations are consistent with the predictions of \citet{bell1995}
for a system where thermal instability is triggered at an intermediate
radius within the metastable disk area. Their model predicts a rapid
ionization from propagating toward the star, causing a short-lived
brightness peak, similar to what we observed in 2010 September. In
parallel, a second, slower ionization front starts to expand outward,
which can be the physical reason behind the re-brightening of
HBC~722 after 2011 September. Our study of HBC\,722 demonstrated that
accretion-related outbursts can occur in young stellar objects even
with very low mass disks. Our results strengthen the theory that
eruptive phenomena may appear throughout star formation from the
embedded phase to the Class\,II phase, and that possibly all young
stars undergo phases of temporarily increased accretion.

\begin{acknowledgements}

We thank the referee for his/her comments that greatly improved the
paper and Z.~Zhu for useful discussions on the accretion disk
fitting. This work was supported by the Momentum grant of the MTA CSFK
Lend\"ulet Disk Research Group, the Lend\"ulet grant LP2012-31 of the
Hungarian Academy of Sciences, and the Hungarian Research Fund OTKA
grant K101393. A.~M.~acknowledges support from the Bolyai Research
Fellowship of the Hungarian Academy of Sciences.

The William Herschel Telescope and its service programme are operated
on the island of La Palma by the Isaac Newton Group in the Spanish
Observatorio del Roque de los Muchachos of the Instituto de
Astrof\'\i{}sica de Canarias.

This work is based in part on observations made with the Telescopio
Carlos Sanchez operated on the island of Tenerife by the Instituto de
Astrof\'\i{}sica de Canarias in the Observatorio del Teide. The
authors wish to thank the telescope manager A.~Oscoz, the support
astronomers and telescope operators for their help during the
observations, as well as the service mode observers.

Based on observations made with the Gran Telescopio Canarias (GTC),
installed in the Spanish Observatorio del Roque de los Muchachos of the
Instituto de Astrof\'\i{}sica de Canarias, in the island of La Palma.

This work is partly based on observations carried out under project
number VA6C with the IRAM Plateau de Bure Interferometer and under
project number 260-11 with the IRAM 30\,m Telescope. IRAM is supported
by INSU/CNRS (France), MPG (Germany) and IGN (Spain). This work has
benefited from research funding from the European Community's sixth
Framework Programme under RadioNet R113CT 2003 5058187.

\end{acknowledgements}

\bibliographystyle{aa}
\bibliography{paper}{}

\begin{thebibliography}{40}
\expandafter\ifx\csname natexlab\endcsname\relax\def\natexlab#1{#1}\fi

\bibitem[{{Acosta-Pulido} {et~al.}(2007){Acosta-Pulido}, {Kun},
  {{\'A}brah{\'a}m}, {K{\'o}sp{\'a}l}, {Csizmadia}, {Kiss}, {Mo{\'o}r},
  {Szabados}, {Benk{\H o}}, {Barrena Delgado}, {Charcos-Llorens}, {Eredics},
  {Kiss}, {Manchado}, {R{\'a}cz}, {Ramos Almeida}, {Sz{\'e}kely}, \&
  {Vidal-N{\'u}{\~n}ez}}]{acosta2007}
{Acosta-Pulido}, J.~A., {Kun}, M., {{\'A}brah{\'a}m}, P., {et~al.} 2007, \aj,
  133, 2020

\bibitem[{{Antoniucci} {et~al.}(2013){Antoniucci}, {Arkharov}, {Klimanov},
  {Lorenzetti}, {Giannini}, {Di Paola}, \& {Larionov}}]{antoniucci2013}
{Antoniucci}, S., {Arkharov}, A., {Klimanov}, S., {et~al.} 2013, The
  Astronomer's Telegram, 5023

\bibitem[{{Armond} {et~al.}(2011){Armond}, {Reipurth}, {Bally}, \&
  {Aspin}}]{armond2011}
{Armond}, T., {Reipurth}, B., {Bally}, J., \& {Aspin}, C. 2011, \aap, 528, A125

\bibitem[{{Aspin} \& {Reipurth}(2009)}]{aspin2009}
{Aspin}, C. \& {Reipurth}, B. 2009, \aj, 138, 1137

\bibitem[{{Audard} {et~al.}(2014){Audard}, {{\'A}brah{\'a}m}, {Dunham},
  {Green}, {Grosso}, {Hamaguchi}, {Kastner}, {K{\'o}sp{\'a}l}, {Lodato},
  {Romanova}, {Skinner}, {Vorobyov}, \& {Zhu}}]{audard2014}
{Audard}, M., {{\'A}brah{\'a}m}, P., {Dunham}, M.~M., {et~al.} 2014, ArXiv
  e-prints

\bibitem[{{Baek} {et~al.}(2015){Baek}, {Pak}, {Green}, {Meschiari}, {Lee},
  {Jeon}, {Choi}, {Im}, {Sung}, \& {Park}}]{baek2015}
{Baek}, G., {Pak}, S., {Green}, J.~D., {et~al.} 2015, \aj, 149, 73

\bibitem[{{Bell} \& {Lin}(1994)}]{bell1994}
{Bell}, K.~R. \& {Lin}, D.~N.~C. 1994, \apj, 427, 987

\bibitem[{{Bell} {et~al.}(1995){Bell}, {Lin}, {Hartmann}, \&
  {Kenyon}}]{bell1995}
{Bell}, K.~R., {Lin}, D.~N.~C., {Hartmann}, L.~W., \& {Kenyon}, S.~J. 1995,
  \apj, 444, 376

\bibitem[{{Castelli} \& {Kurucz}(2004)}]{kurucz2004}
{Castelli}, F. \& {Kurucz}, R.~L. 2004, ArXiv Astrophysics e-prints

\bibitem[{{Cepa}(2010)}]{Cepa10}
{Cepa}, J. 2010, in Highlights of Spanish Astrophysics V, ed. J.~M. {Diego},
  L.~J. {Goicoechea}, J.~I. {Gonz{\'a}lez-Serrano}, \& J.~{Gorgas}, 15

\bibitem[{{Cepa} {et~al.}(2003){Cepa}, {Aguiar-Gonzalez}, {Bland-Hawthorn},
  {Castaneda}, {Cobos}, {Correa}, {Espejo}, {Fragoso-Lopez}, {Fuentes},
  {Gigante}, {Gonzalez}, {Gonzalez-Escalera}, {Gonzalez-Serrano},
  {Joven-Alvarez}, {Lopez-Ruiz}, {Militello}, {Cano}, {Perez}, {Perez},
  {Rasilla}, {Sanchez}, \& {Tejada}}]{Cepa03}
{Cepa}, J., {Aguiar-Gonzalez}, M., {Bland-Hawthorn}, J., {et~al.} 2003, in
  Society of Photo-Optical Instrumentation Engineers (SPIE) Conference Series,
  Vol. 4841, Instrument Design and Performance for Optical/Infrared
  Ground-based Telescopes, ed. M.~{Iye} \& A.~F.~M. {Moorwood}, 1739--1749

\bibitem[{{Cohen} \& {Kuhi}(1979)}]{cohen1979}
{Cohen}, M. \& {Kuhi}, L.~V. 1979, \apjs, 41, 743

\bibitem[{{Dunham} {et~al.}(2012){Dunham}, {Arce}, {Bourke}, {Chen}, {van
  Kempen}, \& {Green}}]{dunham2012b}
{Dunham}, M.~M., {Arce}, H.~G., {Bourke}, T.~L., {et~al.} 2012, \apj, 755, 157

\bibitem[{{Gramajo} {et~al.}(2014){Gramajo}, {Rod{\'o}n}, \&
  {G{\'o}mez}}]{gramajo2014}
{Gramajo}, L.~V., {Rod{\'o}n}, J.~A., \& {G{\'o}mez}, M. 2014, \aj, 147, 140

\bibitem[{{Gratier} {et~al.}(2013){Gratier}, {Pety}, {Guzm{\'a}n}, {Gerin},
  {Goicoechea}, {Roueff}, \& {Faure}}]{gratier2013}
{Gratier}, P., {Pety}, J., {Guzm{\'a}n}, V., {et~al.} 2013, \aap, 557, A101

\bibitem[{{Green} {et~al.}(2013{\natexlab{a}}){Green}, {Evans},
  {K{\'o}sp{\'a}l}, {Herczeg}, {Quanz}, {Henning}, {van Kempen}, {Lee},
  {Dunham}, {Meeus}, {Bouwman}, {Chen}, {G{\"u}del}, {Skinner}, {Liebhart}, \&
  {Merello}}]{green2013}
{Green}, J.~D., {Evans}, II, N.~J., {K{\'o}sp{\'a}l}, {\'A}., {et~al.}
  2013{\natexlab{a}}, \apj, 772, 117

\bibitem[{{Green} {et~al.}(2011){Green}, {Evans}, {K{\'o}sp{\'a}l}, {van
  Kempen}, {Herczeg}, {Quanz}, {Henning}, {Lee}, {Dunham}, {Meeus}, {Bouwman},
  {van Dishoeck}, {Chen}, {G{\"u}del}, {Skinner}, {Merello}, {Pooley},
  {Rebull}, \& {Guieu}}]{green2011}
{Green}, J.~D., {Evans}, II, N.~J., {K{\'o}sp{\'a}l}, {\'A}., {et~al.} 2011,
  \apjl, 731, L25

\bibitem[{{Green} {et~al.}(2013{\natexlab{b}}){Green}, {Robertson}, {Baek},
  {Pooley}, {Pak}, {Im}, {Lee}, {Jeon}, {Choi}, \& {Meschiari}}]{green2013b}
{Green}, J.~D., {Robertson}, P., {Baek}, G., {et~al.} 2013{\natexlab{b}}, \apj,
  764, 22

\bibitem[{{Hackstein} {et~al.}(2015){Hackstein}, {Haas}, {K{\'o}sp{\'a}l},
  {Hambsch}, {Chini}, {{\'A}brah{\'a}m}, {Mo{\'o}r}, {Pozo Nu{\~n}ez},
  {Ramolla}, {Westhues}, {Kaderhandt}, {Fein}, {Barr Dom{\'{\i}}nguez}, \&
  {Hodapp}}]{hackstein2015}
{Hackstein}, M., {Haas}, M., {K{\'o}sp{\'a}l}, {\'A}., {et~al.} 2015, \aap,
  582, L12

\bibitem[{{Hartmann} \& {Kenyon}(1996)}]{hk96}
{Hartmann}, L. \& {Kenyon}, S.~J. 1996, \araa, 34, 207

\bibitem[{{Herbig}(1977)}]{herbig77}
{Herbig}, G.~H. 1977, \apj, 217, 693

\bibitem[{{Kenyon} {et~al.}(1990){Kenyon}, {Hartmann}, {Strom}, \&
  {Strom}}]{kenyon1990}
{Kenyon}, S.~J., {Hartmann}, L.~W., {Strom}, K.~M., \& {Strom}, S.~E. 1990,
  \aj, 99, 869

\bibitem[{{K{\'o}sp{\'a}l} {et~al.}(2011){K{\'o}sp{\'a}l}, {{\'A}brah{\'a}m},
  {Acosta-Pulido}, {Ar{\'e}valo Morales}, {Carnerero}, {Elek}, {Kelemen},
  {Kun}, {P{\'a}l}, {Szak{\'a}ts}, \& {Vida}}]{kospal2011}
{K{\'o}sp{\'a}l}, {\'A}., {{\'A}brah{\'a}m}, P., {Acosta-Pulido}, J.~A.,
  {et~al.} 2011, \aap, 527, A133

\bibitem[{{Kun} {et~al.}(2011){Kun}, {Szegedi-Elek}, {Mo{\'o}r},
  {K{\'o}sp{\'a}l}, {{\'A}brah{\'a}m}, {Apai}, {Kiss}, {Klagyivik}, {Magakian},
  {Mez{\H o}}, {Movsessian}, {P{\'a}l}, {R{\'a}cz}, \& {Rogers}}]{kun2011}
{Kun}, M., {Szegedi-Elek}, E., {Mo{\'o}r}, A., {et~al.} 2011, \mnras, 413, 2689

\bibitem[{{Lee} {et~al.}(2015){Lee}, {Park}, {Green}, {Cochran}, {Kang}, {Lee},
  \& {Sung}}]{lee2015}
{Lee}, J.-E., {Park}, S., {Green}, J.~D., {et~al.} 2015, \apj, 807, 84

\bibitem[{{Lequeux}(2005)}]{lequeux2005}
{Lequeux}, J. 2005, {The Interstellar Medium}

\bibitem[{{Liebhart} {et~al.}(2014){Liebhart}, {G{\"u}del}, {Skinner}, \&
  {Green}}]{liebhart2014}
{Liebhart}, A., {G{\"u}del}, M., {Skinner}, S.~L., \& {Green}, J. 2014, \aap,
  570, L11

\bibitem[{{Lorenzetti} {et~al.}(2012){Lorenzetti}, {Efimova}, {Larionov},
  {Arkharov}, {Gorshanov}, {Giannini}, {Antoniucci}, \& {Di
  Paola}}]{lorenzetti2012}
{Lorenzetti}, D., {Efimova}, N., {Larionov}, V., {et~al.} 2012, The
  Astronomer's Telegram, 4123, 1

\bibitem[{{Miller} {et~al.}(2011){Miller}, {Hillenbrand}, {Covey}, {Poznanski},
  {Silverman}, {Kleiser}, {Rojas-Ayala}, {Muirhead}, {Cenko}, {Bloom},
  {Kasliwal}, {Filippenko}, {Law}, {Ofek}, {Dekany}, {Rahmer}, {Hale}, {Smith},
  {Quimby}, {Nugent}, {Jacobsen}, {Zolkower}, {Velur}, {Walters}, {Henning},
  {Bui}, {McKenna}, {Kulkarni}, {Klein}, {Kandrashoff}, \&
  {Morton}}]{miller2011}
{Miller}, A.~A., {Hillenbrand}, L.~A., {Covey}, K.~R., {et~al.} 2011, \apj,
  730, 80

\bibitem[{{Rebull} {et~al.}(2011){Rebull}, {Guieu}, {Stauffer}, {Hillenbrand},
  {Noriega-Crespo}, {Stapelfeldt}, {Carey}, {Carpenter}, {Cole}, {Padgett},
  {Strom}, \& {Wolff}}]{rebull2011}
{Rebull}, L.~M., {Guieu}, S., {Stauffer}, J.~R., {et~al.} 2011, \apjs, 193, 25

\bibitem[{{Ricci} {et~al.}(2010){Ricci}, {Testi}, {Natta}, \&
  {Brooks}}]{ricci2010}
{Ricci}, L., {Testi}, L., {Natta}, A., \& {Brooks}, K.~J. 2010, \aap, 521, A66

\bibitem[{{Semkov} \& {Peneva}(2010)}]{semkov2010a}
{Semkov}, E. \& {Peneva}, S. 2010, The Astronomer's Telegram, 2801, 1

\bibitem[{{Semkov} {et~al.}(2014){Semkov}, {Peneva}, {Ibryamov}, \&
  {Dimitrov}}]{semkov2014}
{Semkov}, E.~H., {Peneva}, S.~P., {Ibryamov}, S.~I., \& {Dimitrov}, D.~P. 2014,
  Bulgarian Astronomical Journal, 20, 59

\bibitem[{{Semkov} {et~al.}(2010){Semkov}, {Peneva}, {Munari}, {Milani}, \&
  {Valisa}}]{semkov2010}
{Semkov}, E.~H., {Peneva}, S.~P., {Munari}, U., {Milani}, A., \& {Valisa}, P.
  2010, \aap, 523, L3

\bibitem[{{Semkov} {et~al.}(2012){Semkov}, {Peneva}, {Munari}, {Tsvetkov},
  {Jurdana-{\v S}epi{\'c}}, {de Miguel}, {Schwartz}, {Dimitrov}, {Kjurkchieva},
  \& {Radeva}}]{semkov2012}
{Semkov}, E.~H., {Peneva}, S.~P., {Munari}, U., {et~al.} 2012, \aap, 542, A43

\bibitem[{{Siess} {et~al.}(2000){Siess}, {Dufour}, \& {Forestini}}]{siess2000}
{Siess}, L., {Dufour}, E., \& {Forestini}, M. 2000, \aap, 358, 593

\bibitem[{{Sung} {et~al.}(2013){Sung}, {Park}, {Yang}, {Lee}, {Yoon}, {Lee},
  {Kang}, {Park}, {Cho}, \& {Park}}]{sung2013}
{Sung}, H.-I., {Park}, W.-K., {Yang}, Y., {et~al.} 2013, Journal of Korean
  Astronomical Society, 46, 253

\bibitem[{{Wilson} \& {Rood}(1994)}]{wilson1994}
{Wilson}, T.~L. \& {Rood}, R. 1994, \araa, 32, 191

\bibitem[{{Wright} {et~al.}(2010){Wright}, {Eisenhardt}, {Mainzer}, {Ressler},
  {Cutri}, {Jarrett}, {Kirkpatrick}, {Padgett}, {McMillan}, {Skrutskie},
  {Stanford}, {Cohen}, {Walker}, {Mather}, {Leisawitz}, {Gautier}, {McLean},
  {Benford}, {Lonsdale}, {Blain}, {Mendez}, {Irace}, {Duval}, {Liu}, {Royer},
  {Heinrichsen}, {Howard}, {Shannon}, {Kendall}, {Walsh}, {Larsen}, {Cardon},
  {Schick}, {Schwalm}, {Abid}, {Fabinsky}, {Naes}, \& {Tsai}}]{wright2010}
{Wright}, E.~L., {Eisenhardt}, P.~R.~M., {Mainzer}, A.~K., {et~al.} 2010, \aj,
  140, 1868

\bibitem[{{Zhu} {et~al.}(2007){Zhu}, {Hartmann}, {Calvet}, {Hernandez},
  {Muzerolle}, \& {Tannirkulam}}]{zhu2007}
{Zhu}, Z., {Hartmann}, L., {Calvet}, N., {et~al.} 2007, \apj, 669, 483

\end{thebibliography}


\Online
\vspace*{70mm}
\hspace*{65mm}
\mbox{\LARGE{Online Material}}

\longtab{1}{
\begin{longtable}{cccccccccc}
\caption{\label{tab:phot} Optical and near-IR photometry in magnitudes for
  HBC\,722.}\\
\hline\hline
Date & JD$\,{-}\,$2,400,000 & $B$ & $V$ & $R$    & $I$     & $J$      & $H$      & $K_S$    & Telescope\\
\hline
\endfirsthead
\caption{continued.}\\
\hline\hline
Date & JD$\,{-}\,$2,400,000 & $B$ & $V$ & $R$    & $I$     & $J$      & $H$      & $K_S$    & Telescope\\
\hline
\endhead
\hline
\endfoot

2010-12-03 & 55534.34 &          &          &          &          & 10.36(1) & 9.41(1)  & 8.90(1)  & TCS     \\
2011-03-24 & 55644.63 & 16.58(3) & 14.95(2) & 13.87(2) & 12.68(1) &          &          &          & Schmidt \\
2011-03-25 & 55645.56 &          & 14.92(2) & 13.83(1) & 12.63(1) &          &          &          & RCC     \\
2011-03-30 & 55650.59 &          & 15.01(1) & 13.85(1) & 12.66(1) &          &          &          & RCC     \\
2011-03-31 & 55651.58 & 16.74(6) & 14.99(2) & 13.87(1) & 12.65(1) &          &          &          & RCC     \\
2011-04-10 & 55661.55 & 16.48(5) & 14.93(5) & 13.82(1) & 12.66(2) &          &          &          & Schmidt \\
2011-04-23 & 55674.55 &          & 15.09(1) & 13.91(1) & 12.70(1) &          &          &          & RCC     \\
2011-04-27 & 55678.65 &          & 15.00(1) & 13.85(1) &          &          &          &          & IAC80   \\
2011-04-27 & 55678.70 &          &          &          &          & 10.72(1) & 9.70(1)  & 9.11(1)  & TCS     \\
2011-05-26 & 55707.52 &          & 15.03(1) & 13.87(1) & 12.66(1) &          &          &          & RCC     \\
2011-06-05 & 55717.52 &          & 15.17(1) & 13.98(1) & 12.65(2) &          &          &          & RCC     \\
2011-06-19 & 55732.38 & 16.86(19)& 14.99(2) & 13.85(1) & 12.62(4) &          &          &          & Schmidt \\
2011-06-20 & 55733.42 & 16.91(22)& 14.97(8) & 13.78(1) & 12.61(2) &          &          &          & Schmidt \\
2011-06-26 & 55739.48 & 16.55(4) & 14.85(2) & 13.68(1) & 12.47(1) &          &          &          & RCC     \\
2011-07-21 & 55763.62 &          &          &          &          & 10.65(2) & 9.67(3)  & 9.17(1)  & WHT/LIRIS \\
2011-08-04 & 55777.61 &          &          &          &          & 10.60(1) & 9.60(1)  & 9.03(1)  & TCS     \\
2011-08-05 & 55779.34 & 16.71(4) & 14.95(1) & 13.81(1) & 12.60(2) &          &          &          & Schmidt \\
2011-08-09 & 55782.63 &          &          &          &          & 10.57(1) & 9.59(1)  & 9.00(1)  & TCS     \\
2011-08-11 & 55785.38 & 16.75(5) & 15.03(2) & 13.87(3) & 12.64(1) &          &          &          & Schmidt \\
2011-08-12 & 55786.36 & 16.35(17)& 14.94(7) & 13.80(1) & 12.60(2) &          &          &          & Schmidt \\
2011-08-14 & 55788.33 & 16.86(27)& 14.91(19)& 13.79(4) & 12.56(3) &          &          &          & Schmidt \\
2011-08-15 & 55789.31 & 16.66(12)& 14.98(3) & 13.75(2) & 12.57(8) &          &          &          & Schmidt \\
2011-08-15 & 55789.43 &          &          &          &          & 10.55(3) & 9.60(1)  & 9.00(1)  & TCS     \\
2011-08-17 & 55791.41 & 16.69(8) & 14.91(1) & 13.73(1) & 12.54(1) &          &          &          & Schmidt \\
2011-08-19 & 55792.56 & 16.73(5) & 14.89(4) & 13.74(5) & 12.59(1) &          &          &          & Schmidt \\
2011-08-20 & 55794.49 & 16.65(3) & 14.91(2) & 13.74(2) & 12.57(2) &          &          &          & Schmidt \\
2011-08-24 & 55797.57 & 16.59(17)& 14.89(3) & 13.71(1) & 12.53(1) &          &          &          & Schmidt \\
2011-08-25 & 55799.36 & 16.74(04)& 14.98(1) & 13.79(1) & 12.57(1) &          &          &          & RCC     \\
2011-08-29 & 55803.37 & 16.67(9) & 14.98(1) & 13.81(1) & 12.59(1) &          &          &          & RCC     \\
2011-08-30 & 55804.39 & 16.71(2) & 14.95(1) & 13.79(1) & 12.58(1) &          &          &          & RCC     \\
2011-09-03 & 55807.51 & 16.60(5) & 15.02(1) & 13.85(1) & 12.67(1) &          &          &          & Schmidt \\
2011-09-06 & 55811.30 & 16.78(6) & 15.04(3) & 13.90(1) & 12.65(1) &          &          &          & Schmidt \\
2011-09-09 & 55813.63 &          &          &          &          & 10.62(1) & 9.60(1)  & 9.01(1)  & TCS     \\
2011-09-09 & 55814.36 & 17.07(20)& 15.07(7) & 13.98(12)& 12.73(4) &          &          &          & Schmidt \\
2011-09-09 & 55814.43 &          &          &          &          & 10.60(1) & 9.61(1)  & 9.04(2)  & TCS     \\
2011-09-11 & 55816.43 &          &          &          &          & 10.59(1) & 9.59(1)  & 9.04(2)  & TCS     \\
2011-09-12 & 55817.46 &          &          &          &          & 10.60(1) & 9.61(1)  & 9.03(1)  & TCS     \\
2011-09-13 & 55818.53 &          &          &          &          & 10.61(1) & 9.56(1)  & 9.00(1)  & TCS     \\
2011-09-16 & 55821.33 & 16.66(4) & 15.00(2) & 13.78(2) & 12.51(1) &          &          &          & Schmidt \\
2011-09-17 & 55822.32 & 16.75(5) & 14.94(4) & 13.77(2) & 12.54(3) &          &          &          & Schmidt \\
2011-09-21 & 55826.34 & 16.56(2) & 14.92(3) & 13.77(1) & 12.57(1) &          &          &          & Schmidt \\
2011-09-24 & 55829.32 & 16.75(3) & 14.96(1) & 13.78(1) & 12.57(1) &          &          &          & RCC     \\
2011-09-25 & 55830.28 & 16.53(12)& 14.80(3) & 13.75(4) & 12.47(2) &          &          &          & Schmidt \\
2011-09-25 & 55830.33 & 16.74(2) & 14.95(1) & 13.77(1) & 12.53(1) &          &          &          & RCC     \\
2011-09-26 & 55831.33 & 16.80(2) & 15.01(1) & 13.81(1) & 12.53(1) &          &          &          & RCC     \\
2011-10-07 & 55842.41 &          &          &          &          & 10.50(1) & 9.51(1)  & 8.94(1)  & TCS     \\
2011-10-08 & 55843.41 &          &          &          &          & 10.50(1) & 9.51(1)  & 8.92(2)  & TCS     \\
2011-10-11 & 55846.44 &          &          &          &          & 10.46(1) & 9.52(1)  & 9.00(2)  & TCS     \\
2011-10-26 & 55861.24 & 16.46(19)& 14.80(8) & 13.60(3) & 12.43(2) &          &          &          & Schmidt \\
2011-10-27 & 55862.28 & 16.54(8) & 14.62(16)& 13.68(2) & 12.38(6) &          &          &          & Schmidt \\
2011-10-27 & 55862.46 &          &          &          &          & 10.42(1) & 9.44(2)  & 8.84(3)  & TCS     \\
2011-10-28 & 55863.20 & 16.34(5) & 14.69(2) & 13.56(1) & 12.41(1) &          &          &          & Schmidt \\
2011-10-29 & 55864.42 & 16.32(7) & 14.71(5) & 13.51(2) & 12.35(1) &          &          &          & Schmidt \\
2011-10-30 & 55865.40 & 16.36(2) & 14.64(1) & 13.55(1) & 12.40(1) &          &          &          & Schmidt \\
2011-10-31 & 55866.40 & 16.39(3) & 14.62(2) & 13.53(1) & 12.37(2) &          &          &          & Schmidt \\
2011-10-31 & 55866.40 &          &          &          &          & 10.39(1) & 9.43(1)  & 8.89(1)  & TCS     \\
2011-11-01 & 55867.23 & 16.40(8) & 14.71(2) & 13.54(1) & 12.40(2) &          &          &          & Schmidt \\
2011-11-01 & 55867.39 &          &          &          &          & 10.41(1) & 9.41(1)  & 8.89(1)  & TCS     \\
2011-11-02 & 55868.22 & 16.41(2) & 14.68(1) & 13.52(1) & 12.38(2) &          &          &          & Schmidt \\
2011-11-02 & 55868.37 &          &          &          &          & 10.42(1) & 9.45(1)  & 8.90(1)  & TCS     \\
2011-11-03 & 55869.39 & 16.30(11)& 14.61(22)& 13.66(10)& 12.50(8) &          &          &          & Schmidt \\
2011-11-04 & 55870.39 &          &          &          &          & 10.40(1) & 9.44(1)  & 8.90(1)  & TCS     \\
2011-11-05 & 55871.37 &          &          &          &          & 10.41(1) & 9.46(1)  & 8.89(1)  & TCS     \\
2011-11-06 & 55872.42 &          &          &          &          & 10.38(1) & 9.44(1)  & 8.90(1)  & TCS     \\
2011-11-08 & 55874.38 &          &          &          &          & 10.35(1) & 9.42(1)  & 8.90(1)  & TCS     \\
2011-11-13 & 55879.26 &          & 14.69(1) & 13.51(3) & 12.35(1) &          &          &          & RCC     \\
2011-11-22 & 55888.19 & 16.22(19)& 14.60(3) & 13.44(3) & 12.32(3) &          &          &          & Schmidt \\
2011-11-23 & 55889.18 & 16.23(5) & 14.60(1) & 13.51(2) & 12.32(1) &          &          &          & Schmidt \\
2011-11-24 & 55890.18 & 16.25(2) & 14.56(3) & 13.45(1) & 12.28(1) &          &          &          & Schmidt \\
2011-11-25 & 55891.22 & 16.22(8) & 14.58(3) & 13.46(3) & 12.32(2) &          &          &          & Schmidt \\
2011-11-27 & 55893.28 & 16.23(3) & 14.61(1) & 13.46(2) & 12.32(1) &          &          &          & Schmidt \\
2011-11-28 & 55894.18 & 16.29(1) & 14.64(4) & 13.45(1) & 12.31(1) &          &          &          & Schmidt \\
2011-11-29 & 55895.18 & 16.34(8) & 14.72(1) & 13.49(1) & 12.32(2) &          &          &          & Schmidt \\
2011-11-29 & 55895.32 &          &          &          &          & 10.31(5) & 9.35(1)  & 8.83(1)  & TCS     \\
2011-11-30 & 55896.18 & 16.20(9) & 14.57(1) & 13.41(2) & 12.22(1) &          &          &          & Schmidt \\
2011-11-30 & 55896.32 &          &          &          &          & 10.30(1) & 9.34(1)  & 8.84(1)  & TCS     \\
2011-12-31 & 55927.18 & 16.09(7) & 14.31(7) & 13.24(1) & 12.09(1) &          &          &          & Schmidt \\
2012-01-05 & 55932.32 & 16.01(2) & 14.29(1) & 13.13(1) & 12.01(1) &          &          &          & IAC80   \\
2012-01-12 & 55939.31 &          &          &          &          & 10.10(1) & 9.16(1)  & 8.63(1)  & TCS     \\
2012-01-12 & 55939.31 &          &          &          & 11.95(1) &          &          &          & IAC80   \\
2012-02-10 & 55968.22 & 16.12(14)& 14.26(5) & 13.01(3) & 11.89(2) &          &          &          & Schmidt \\
2012-04-28 & 56045.59 & 15.88(7) & 14.04(6) & 12.97(2) & 11.84(1) &          &          &          & Schmidt \\
2012-05-20 & 56067.51 & 15.60(4) & 14.01(1) & 12.90(1) & 11.81(1) &          &          &          & RCC     \\
2012-05-24 & 56071.51 & 16.04(13)& 14.02(1) & 12.91(1) & 11.83(1) &          &          &          & RCC     \\
2012-06-01 & 56079.52 & 15.57(8) & 13.92(5) & 12.97(2) & 11.87(3) &          &          &          & Schmidt \\
2012-06-07 & 56086.46 & 15.72(10)& 14.01(5) & 12.95(3) & 11.81(2) &          &          &          & Schmidt \\
2012-06-19 & 56098.49 & 15.64(7) & 13.95(8) & 12.77(2) & 11.72(1) &          &          &          & Schmidt \\
2012-07-22 & 56131.32 & 15.86(24)& 13.89(10)& 12.85(2) & 11.70(1) &          &          &          & Schmidt \\
2012-07-23 & 56132.32 & 15.23(25)& 13.94(7) & 12.79(1) & 11.60(8) &          &          &          & Schmidt \\
2012-07-25 & 56134.41 & 15.78(5) & 13.85(2) & 12.72(1) & 11.57(1) &          &          &          & IAC80   \\
2012-07-25 & 56134.58 &          &          &          &          & 9.82(1)  & 8.92(1)  & 8.38(2)  & TCS     \\
2012-08-09 & 56149.37 & 15.43(1) & 13.83(1) & 12.72(1) & 11.63(1) &          &          &          & RCC     \\
2012-08-19 & 56159.40 & 15.33(2) & 13.74(1) & 12.64(1) & 11.57(1) &          &          &          & RCC     \\
2012-08-20 & 56160.41 & 16.31(5) & 13.99(3) & 12.80(4) & 11.62(1) &          &          &          & IAC80   \\
2012-08-20 & 56160.70 &          &          &          &          & 9.80(2)  & 8.84(1)  & 8.36(1)  & TCS     \\
2012-08-25 & 56164.37 & 15.43(4) & 13.79(3) & 12.71(2) & 11.61(2) &          &          &          & RCC     \\
2012-08-26 & 56165.33 & 15.14(32)& 13.84(2) & 12.71(2) & 11.67(2) &          &          &          & RCC     \\
2012-09-16 & 56187.36 &          & 13.73(1) & 12.66(1) & 11.61(2) &          &          &          & RCC     \\
2012-10-09 & 56210.39 &          &          &          &          & 9.82(2)  & 8.85(1)  & 8.26(1)  & TCS     \\
2012-10-10 & 56211.44 &          &          &          &          & 9.78(1)  & 8.81(1)  & 8.25(1)  & TCS     \\
2012-10-11 & 56212.43 &          &          &          &          & 9.73(1)  & 8.79(1)  & 8.21(1)  & TCS     \\
2012-10-13 & 56214.43 &          &          &          &          & 9.71(1)  & 8.75(1)  & 8.23(1)  & TCS     \\
2013-07-24 & 56497.54 &          & 13.56(5) & 12.27(2) & 11.24(2) &          &          &          & Schmidt \\
2013-08-08 & 56513.36 & 14.83(3) & 13.30(1) & 12.26(1) & 11.18(2) &          &          &          & Schmidt \\
2013-08-17 & 56522.41 & 14.94(7) & 13.38(3) & 12.30(1) & 11.25(1) &          &          &          & Schmidt \\
2013-08-31 & 56536.34 & 15.24(20)& 13.39(2) & 12.31(1) & 11.21(1) &          &          &          & RCC     \\
2013-09-01 & 56537.41 & 15.05(3) & 13.42(8) & 12.32(1) & 11.25(1) &          &          &          & RCC     \\
2013-09-03 & 56539.45 & 15.08(20)& 13.37(3) & 12.34(1) & 11.25(1) &          &          &          & RCC     \\
2013-09-04 & 56540.33 & 15.30(20)& 13.61(18)& 12.33(4) & 11.26(2) &          &          &          & RCC     \\
2014-02-24 & 56712.66 & 15.03(4) & 13.38(4) & 12.33(3) & 11.29(3) &          &          &          & Schmidt \\
2014-04-08 & 56755.61 & 14.99(6) & 13.33(5) & 12.27(2) & 11.24(3) &          &          &          & Schmidt \\
2014-04-30 & 56777.58 & 15.08(6) & 13.45(3) & 12.32(4) & 11.24(4) &          &          &          & Schmidt \\
2014-06-05 & 56813.53 & 15.02(7) & 13.53(3) & 12.41(3) & 11.34(3) &          &          &          & Schmidt \\
2015-01-06 & 57029.22 & 14.98(6) & 13.42(5) & 12.26(3) & 11.18(3) &          &          &          & Schmidt \\
2015-02-17 & 57070.68 & 15.21(14)& 13.36(5) & 12.34(2) & 11.16(2) &          &          &          & Schmidt \\
2015-02-18 & 57071.68 & 15.12(6) & 13.37(5) & 12.27(2) & 11.20(2) &          &          &          & Schmidt \\
2015-03-17 & 57098.64 & 15.00(4) & 13.30(5) & 12.25(1) & 11.13(2) &          &          &          & Schmidt \\
2015-04-09 & 57121.62 & 14.88(19)& 13.57(7) & 12.28(1) & 11.19(3) &          &          &          & Schmidt \\
2015-04-13 & 57125.59 & 14.99(6) & 13.35(2) & 12.33(4) & 11.26(3) &          &          &          & Schmidt \\
2015-04-20 & 57132.53 & 14.87(16)& 13.38(3) & 12.30(2) & 11.22(2) &          &          &          & Schmidt \\
2015-04-23 & 57135.57 & 15.07(5) & 13.39(5) & 12.26(1) & 11.20(2) &          &          &          & Schmidt \\
2015-05-08 & 57150.51 & 15.16(8) & 13.58(3) & 12.41(3) & 11.28(4) &          &          &          & Schmidt \\
2015-05-16 & 57159.44 & 15.12(14)& 13.33(9) & 12.31(2) & 11.25(4) &          &          &          & Schmidt \\
2015-05-20 & 57163.72 &          &          &          &          & 9.30(1)  & 8.35(1)  & 7.79(1)  & TCS     \\
2015-05-21 & 57164.78 &          &          &          &          & 9.30(1)  & 8.39(3)  & 7.81(16) & TCS     \\
2015-06-01 & 57174.54 & 15.03(10)& 13.34(2) & 12.27(1) & 11.20(1) &          &          &          & Schmidt \\
2015-06-01 & 57175.46 & 15.01(10)& 13.38(4) & 12.26(3) & 11.18(2) &          &          &          & Schmidt \\
2015-06-02 & 57176.49 & 15.06(3) & 13.33(4) & 12.28(1) & 11.21(2) &          &          &          & Schmidt \\
2015-06-03 & 57177.44 & 15.08(10)& 13.51(7) & 12.30(1) & 11.20(1) &          &          &          & Schmidt \\
2015-06-06 & 57180.49 & 14.96(8) & 13.41(3) & 12.32(3) & 11.21(4) &          &          &          & Schmidt \\
2015-06-07 & 57181.48 & 15.15(10)& 13.46(6) & 12.31(4) & 11.23(2) &          &          &          & Schmidt \\
2015-06-12 & 57186.45 & 14.83(8) & 13.35(1) & 12.26(3) & 11.18(1) &          &          &          & Schmidt \\
2015-06-18 & 57191.52 & 15.10(10)& 13.33(2) & 12.27(1) & 11.16(1) &          &          &          & Schmidt \\
2015-07-16 & 57220.49 & 14.90(4) & 13.31(2) & 12.21(1) & 11.15(1) &          &          &          & Schmidt \\
2015-07-22 & 57226.43 & 14.89(3) & 13.42(2) & 12.35(1) & 11.26(3) &          &          &          & Schmidt \\
2015-08-05 & 57240.36 & 14.93(2) & 13.39(2) & 12.33(1) & 11.24(1) &          &          &          & Schmidt \\
2015-08-12 & 57247.32 & 14.98(13)& 13.30(6) & 12.28(4) & 11.16(3) &          &          &          & Schmidt \\
2015-08-31 & 57266.37 & 14.91(22)& 13.42(11)& 12.34(7) & 11.23(5) &          &          &          & Schmidt \\
2015-09-01 & 57267.33 & 15.12(35)& 13.40(1) & 12.30(3) & 11.24(1) &          &          &          & Schmidt \\
2015-09-02 & 57268.32 & 15.03(49)& 13.35(4) & 12.26(1) & 11.21(2) &          &          &          & Schmidt \\
2015-09-10 & 57275.56 & 15.02(30)& 13.30(4) & 12.21(3) & 11.06(2) &          &          &          & Schmidt \\
2015-09-15 & 57281.33 &          & 13.37(2) & 12.27(4) & 11.21(2) &          &          &          & Schmidt \\
2015-09-17 & 57283.25 &          & 13.36(4) & 12.31(3) & 11.22(1) &          &          &          & Schmidt \\
2015-09-18 & 57284.25 &          & 13.42(10)& 12.39(5) & 11.21(3) &          &          &          & Schmidt \\
2015-10-22 & 57318.25 &          & 13.52(4) & 12.45(2) & 11.32(2) &          &          &          & Schmidt \\
2015-10-27 & 57323.32 &          & 13.39(4) & 12.34(2) & 11.23(3) &          &          &          & Schmidt \\
2015-11-05 & 57332.19 &          & 13.44(4) & 12.28(5) & 11.30(4) &          &          &          & Schmidt \\
2015-11-08 & 57335.20 &          & 13.46(2) & 12.42(3) & 11.29(1) &          &          &          & Schmidt \\
2015-11-12 & 57339.31 &          &          & 12.34(1) & 11.22(2) &          &          &          & Schmidt \\
2016-03-10 & 57457.63 & 14.95(9) & 13.30(5) & 12.44(7) & 11.24(2) &          &          &          & Schmidt \\
2016-04-03 & 57481.62 & 15.24(12)& 13.32(5) & 12.30(3) & 11.19(6) &          &          &          & Schmidt \\
2016-04-30 & 57508.56 & 15.05(4) & 13.41(3) & 12.32(1) & 11.20(1) &          &          &          & Schmidt \\
2016-05-07 & 57515.53 & 14.94(6) & 13.30(2) & 12.22(4) & 11.15(2) &          &          &          & Schmidt \\
2016-05-21 & 57530.44 & 15.40(7) & 13.52(2) & 12.32(2) & 11.29(2) &          &          &          & Schmidt \\
2016-05-29 & 57538.49 & 15.18(7) & 13.48(2) & 12.25(4) & 11.29(6) &          &          &          & Schmidt \\
2016-05-31 & 57539.54 & 14.97(13)& 13.45(3) & 12.37(1) & 11.29(1) &          &          &          & Schmidt \\
2016-06-08 & 57548.46 & 14.96(6) & 13.30(2) & 12.27(1) & 11.19(1) &          &          &          & Schmidt \\
2016-06-23 & 57563.46 & 14.96(9) & 13.37(9) & 12.26(3) & 11.20(2) &          &          &          & Schmidt \\
2016-07-06 & 57576.45 & 15.05(4) & 13.43(3) & 12.33(4) & 11.25(1) &          &          &          & Schmidt \\
2016-07-12 & 57581.54 & 15.15(5) & 13.45(1) & 12.34(2) & 11.27(3) &          &          &          & Schmidt \\
\end{longtable}}

\onltab{2}{
\begin{table*}
\caption{Spitzer/IRAC and WISE photometry for HBC\,722. Fluxes are not
  color-corrected.}\label{tab:spitzer}
\begin{tabular}{ccccccc}
\hline \hline
Date        & JD$\,{-}\,$2,400,000 & F$_{3.4}$ (Jy)   & F$_{3.6}$ (Jy)   & F$_{4.5}$ (Jy) & F$_{4.6}$ (Jy) & Telescope      \\
\hline
2010-05-29 & 55346.30             & 0.082$\pm$0.004 &                 &                 & 0.098$\pm$0.003 & WISE \\
2010-11-26 & 55527.10             & 0.168$\pm$0.005 &                 &                 & 0.169$\pm$0.008 & WISE \\
2011-09-08 & 55813.02             &                 & 0.127$\pm$0.004 & 0.112$\pm$0.003 &     & Spitzer \\
2011-09-24 & 55829.49             &                 & 0.129$\pm$0.004 & 0.115$\pm$0.004 &     & Spitzer \\
2011-11-29 & 55895.31             &                 & 0.154$\pm$0.005 & 0.136$\pm$0.004 &     & Spitzer \\
2012-01-06 & 55932.50             &                 & 0.169$\pm$0.005 & 0.152$\pm$0.005 &     & Spitzer \\
2012-01-11 & 55938.10             &                 & 0.173$\pm$0.005 & 0.154$\pm$0.005 &     & Spitzer \\
2012-07-25 & 56134.09             &                 & 0.240$\pm$0.007 & 0.216$\pm$0.006 &     & Spitzer \\
2012-08-20 & 56160.46             &                 & 0.248$\pm$0.008 & 0.226$\pm$0.007 &     & Spitzer \\
2012-09-16 & 56186.75             &                 & 0.269$\pm$0.008 & 0.247$\pm$0.008 &     & Spitzer \\
2012-10-12 & 56212.64             &                 & 0.272$\pm$0.008 & 0.248$\pm$0.007 &     & Spitzer \\
\hline
\end{tabular}
\end{table*}
}

\onltab{3}{
\begin{table*}
\caption{IRAM 2.7\,mm photometry for HBC\,722 and for the millimeter
  sources in its vicinity.}\label{tab:iram}
\begin{tabular}{ccccc}
\hline \hline
Name     & Peak flux (mJy/beam) & Total flux (mJy) & Size ($''$) & P.A. ($^{\circ}$) \\
\hline
HBC\,722 & $<$0.08              & $<$0.08       & $\dots$        & $\dots$         \\
MMS1     & 3.63$\pm$0.11        & 4.53$\pm$0.22 & 1.3$\times$1.1 & 115             \\
MMS2     & 1.44$\pm$0.10        & 1.61$\pm$0.18 & 1.1$\times$0.6 & 100             \\
MMS3     & 1.38$\pm$0.11        & 4.03$\pm$0.40 & 4.9$\times$2.0 & 48              \\
MMS4     & 2.07$\pm$0.21        & 4.34$\pm$0.61 & 3.4$\times$1.6 & 31              \\
MMS5     & 0.83$\pm$0.10        & 2.58$\pm$0.30 & 6.1$\times$1.8 & 83              \\
MMS6     & 0.39$\pm$0.05        & 0.68$\pm$0.13 & 3.4$\times$0.2 & 48              \\
MMS7     & $<$0.08              & $<$0.08       & $\dots$        & $\dots$         \\
\hline
\end{tabular}
\end{table*}
}

\end{document}